\UseRawInputEncoding
\documentclass[pra,twocolumn,showpacs,groupedaddress,superscriptaddress,nofootinbib,floatfix,preprintnumbers,longbibliography]{revtex4-1}

\usepackage{amssymb,amsmath,graphicx,color}
\usepackage{bm}
\usepackage[tight]{subfigure}
\usepackage[export]{adjustbox}
\usepackage{braket}
\usepackage[colorlinks=true,citecolor=blue,linkcolor=red]{hyperref}
\usepackage{multirow}
\usepackage{rotating,booktabs}
\usepackage[verbose]{placeins}

\begin{document}

\title{
Toward simulating quantum field theories with controlled phonon-ion dynamics:
\\
A hybrid analog-digital approach
}

\author{Zohreh Davoudi}
\thanks{Also affiliated with the RIKEN Center for Accelerator-based Sciences, Wako, Japan, during the early stages of this work.}
\affiliation{
Maryland Center for Fundamental Physics and Department of Physics, 
University of Maryland, College Park, MD 20742, USA.}

\author{Norbert M. Linke}
\affiliation{
Joint Quantum Institute and Department of Physics,
University of Maryland, College Park, MD 20742}

\author{Guido Pagano}
\affiliation{
Department of Physics and Astronomy, Rice University, 6100 Main Street, Houston, TX 77005, USA.}

\date{\today}

\preprint{UMD-PP-021-02
}

\begin{abstract}
Quantum field theories are the cornerstones of modern physics, providing relativistic and quantum mechanical descriptions of physical systems at the most fundamental level. Simulating real-time dynamics within these theories remains elusive in classical computing. This provides a unique opportunity for quantum simulators, which hold the promise of revolutionizing our simulation capabilities. Trapped-ion systems are successful quantum-simulator platforms for quantum many-body physics and can operate in digital, or gate-based, and analog modes. Inspired by the progress in proposing and realizing quantum simulations of a number of relativistic quantum field theories using trapped-ion systems, and by the hybrid analog-digital proposals for simulating interacting boson-fermion models, we propose hybrid analog-digital quantum simulations of selected quantum field theories, taking recent developments to the next level. On one hand, the semi-digital nature of this proposal offers more flexibility in engineering generic model interactions compared with a fully-analog approach. On the other hand, encoding the bosonic fields onto the phonon degrees of freedom of the trapped-ion system allows a more efficient usage of simulator resources, and a more natural implementation of intrinsic quantum operations in such platforms. This opens up new ways for simulating complex dynamics of e.g., Abelian and non-Abelian gauge theories, by combining the benefits of digital and analog schemes.
\end{abstract}

\maketitle
\section{Introduction
\label{sec:intro}}
\noindent
Quantum field theories (QFTs) provide the underlying quantum-mechanical descriptions of physical systems, from relativistic gauge field theories of the Standard Model of particle physics~\cite{hokim1998, schwartz2013}, to emergent low-energy models in condensed-matter systems~\cite{kleinert1989gauge, fradkin2013field}, to effective field theories in hadronic and nuclear physics~\cite{Kaplan:2005es, Hammer:2019poc}. Classical simulation methods have come a long way to describe phenomena emerging from these underlying theories, with notable examples in the realm of lattice gauge theory (LGT) methods applied in strong-interaction physics~\cite{Aoki:2019cca, Davoudi:2020ngi}. Nonetheless, there is a need for new computational strategies to overcome the limitations of the current methods, in order to achieve real-time simulations of matter, and predictions for equilibrium and out-of-equilibrium phenomena arising from strong-interaction dynamics. Hamiltonian simulation of physical systems, naturally enabled via mapping the problem to a quantum simulator or a quantum computer, is an example of such a strategy, with major advances reported in recent years in proposing~\cite{Byrnes:2005qx, IgnacioCirac:2010us, Lamata:2011me, Jordan:2011ne, Jordan:2011ci, Casanova:2011wh, Zohar:2011cw, Tagliacozzo:2012vg, Banerjee:2012pg, Zohar:2012xf, Zohar:2013zla, Jordan:2014tma, Zohar:2014qma, Mezzacapo:2015bra, Zohar:2016wmo, Moosavian:2017tkv, Zache:2018jbt, Bhattacharyya:2018bbv, Stryker:2018efp, Raychowdhury:2018osk, Klco:2018zqz, Bender:2018rdp, Klco:2019xro, Davoudi:2019bhy, Klco:2019yrb, Lamm:2019uyc, Mueller:2019qqj, Lamm:2019bik, Alexandru:2019nsa, Harmalkar:2020mpd, Shaw:2020udc, Kharzeev:2020kgc, Chakraborty:2020uhf, Liu:2020eoa, Kreshchuk:2020dla, Barata:2020jtq, Haase:2020kaj, Paulson:2020zjd, Ji:2020kjk,  Buser:2020cvn, Bender:2020ztu, Ciavarella:2021nmj, Banuls:2019bmf, Dasgupta:2020itb} and implementing~\cite{gerritsma2010quantum, Martinez:2016yna, Kokail:2018eiw, Klco:2018kyo, Lu:2018pjk, Klco:2019evd, Atas:2021ext, Kreshchuk:2020kcz, Bauer:2019qxa, Bauer:2021gup, Gorg:2018xyc, schweizer2019floquet, Mil:2019pbt, Yang:2020yer, Gustafson:2021imb, Rahman:2021yse} simulations of various field theories on the limited existing quantum hardware, and in formulating suitable Hamiltonian descriptions of QFTs and LGTs for resource-efficient and robust quantum simulations~\cite{Jordan:2011ne, Jordan:2011ci, Klco:2018zqz, Barata:2020jtq, Byrnes:2005qx, Zohar:2014qma, Lamm:2019bik, Ji:2020kjk, Kreshchuk:2020dla, Liu:2020eoa, Buser:2020cvn, Haase:2020kaj, Bender:2020ztu, Anishetty:2009nh, mathur2010n, Raychowdhury:2019iki, Mathur:2016cko, Zohar:2019ygc, Davoudi:2020yln, Ciavarella:2021nmj}. The question we investigate here is complementary to the latter effort: Given a Hamiltonian formulation, what are the efficient implementation strategies that maximally take advantage of available hardware capacity and capability? We focus on trapped-ion quantum simulators as one of the leading quantum-hardware technologies. Nonetheless, applications of this strategy to other platforms, such as cavity QED and superconducting circuits, can be established analogously.

QFTs with bosonic fields are extremely common, from scalar field theory descriptions of phase transitions in condensed-matter systems or of the  inflationary phase of the early universe and the Higgs mechanism, to gauge field theory descriptions of the fundamental forces of nature. Due to their infinite-dimensional Hilbert space, even on a single space-time point, a truncation must be imposed on the various bosonic excitations  to contain their dynamics on a finite simulating hardware. Important work has emerged in understanding, qualitatively or quantitatively, the impact of this truncation, from the simple quantum harmonic oscillator~\cite{somma2015quantum, Macridin:2018gdw, Macridin:2018oli} to scalar field theory~\cite{Jordan:2011ci, Klco:2018zqz} and LGTs~\cite{Hackett:2018cel, Ji:2020kjk, Davoudi:2020yln, Ciavarella:2021nmj}. While quantities in the low-energy subspace of the truncated theories exhibit exponential convergence to the exact values in the full theory, as guaranteed by the Shannon-Nyquist sampling theorem~\cite{Macridin:2018gdw, Klco:2018zqz} in the case of scalar field theory, observables in the high-energy sector or those obtained from long-time expectation values exhibit a slow convergence to the exact values. In particular, as shown in Ref.~\cite{Davoudi:2020yln} for the case of the SU(2) LGT coupled to matter in 1+1 D, for a certain threshold on the gauge-field cutoff  quantities enter a scaling region in which the dependence on the cutoff becomes exponential and systematically improvable. However, for high-energy spectra and long-time dynamical quantities, such a threshold is reached for a large value of the cutoff, increasing the simulation-resource requirement needed to reach a fixed accuracy.

The most common encoding of bosonic degrees of freedom to qubits for digital quantum simulation is to convert their occupation number to a binary representation, the qubit-number requirement of which is logarithmic in the cutoff on the highest excitation of the boson. Most importantly for near-term applications, the number of entangling gates required to implement the dynamics associated with boson-boson or boson-fermion interactions grows polynomially with the system size, as a large number of controlled operations must be introduced on the state of the boson register, see e.g., Refs.~\cite{somma2015quantum,Macridin:2018oli, Shaw:2020udc}. This problem is particularly severe in the case of bosonic field theories as, for example, a scalar field in a finite discretized space represents quantum harmonic oscillators, or bosons, associated with each momentum mode, the number of which scales linearly with the volume of the reciprocal lattice. Each harmonic oscillator then requires the allocation of dedicated qubits and the implementation of associated entangling gates. A cost analysis of this kind for a fully-digital implementation is made explicit in this paper and is contrasted with our analog-digital protocol, to be introduced next.

A trapped-ion quantum simulator~\cite{blatt2012quantum, wineland1998experimental, Monroe21} can operate in an analog mode, with which dynamics generated by the Ising and Heisenberg spin Hamiltonians can be studied~\cite{porras2004effective,islam2011onset,schneider2012experimental,jurcevic2014quasiparticle,richerme2014non,zhang2017observation,hess2017non,neyenhuis2017observation,liu2018confined,Davoudi:2019bhy}, or a digital mode, with which universal computations expressed in terms of single- and two-qubit gates can be performed~\cite{cirac1995quantum,molmer1999multiparticle,solano1999deterministic,milburn2000ion,blatt2008entangled,lanyon2011universal,monroe2013scaling,monroe2014large,Martinez:2016yna,debnath2016demonstration,linke2017experimental,figgatt2018parallel,landsman2019verified,nam2019ground,wright2019benchmarking}. Typically in these systems, specific pairs of internal states encode qubits, which are manipulated via ion-laser or ion-microwave interactions. The collective excitations of the motion of ions in the trap, i.e., the phonons, are often used as mediators of the interactions among the qubits, enabling the engineering of effective spin Hamiltonians or in turn entangling operations among two or more qubits~\cite{Monroe21}. In recent years, promising progress has been reported in controlling the phonon dynamics in a trapped-ion simulator, leading to experimental demonstrations of phonon hopping, phonon interference, and a phonon quantum walk in Paul traps~\cite{Haze12, Toyoda2015, debnath2018observation,Ohira19, tamura2020quantum}. This opens up the possibility of taking advantage of phonons, in addition to the qubits, as dynamical degrees of freedom to encode and process information~\cite{Fluhmann2019,Chen21}. In fact, there exist quantum-simulation proposals for an Abelian LGT, i.e., the lattice Schwinger model,  which take advantage of the phonons in a fully-analog setting to encode the gauge or fermion degrees of freedom~\cite{yang2016analog}, but they are not experimentally feasible yet. A more viable option for near-term applications is to combine the flexibility of gate-based digital simulations with the versatility of both the controllable spin and phonon degrees of freedom in a trapped-ion simulator~\cite{Casanova:2011wh, Casanova:2012zz, Lamata:2013sta, mezzacapo2012digital}. This idea has led to concrete gate-based protocols for simulating interacting fermion-boson models, such as the Holstein model of electron-phonon dynamics in condensed-matter physics~\cite{mezzacapo2012digital}, as well as the first experimental implementation of a one-site boson-fermion dynamics~\cite{Zhang:2018nrx} considering only a single excitation of the boson. In the present work, the ideas in these proposals are taken to the next level to demonstrate that quantum simulation of relativistic QFTs, including LGTs, can benefit from a similar hybrid protocol. 

This work introduces a complete and enhanced set of conventional as well as phonon-based gates, including but not limited to those introduced in Ref.~\cite{mezzacapo2012digital}, and deploys them to construct concrete circuits for the digitized time-evolution operator within a Yukawa theory of scalar-fermion interactions and the $U(1)$ LGT coupled to matter. The gates include single-spin (qubit) operations, and entangling spin-spin, spin-phonon, and phonon-phonon operations. In order to ensure efficient implementation of all these gates in one experiment, normal modes of motion, local modes of motion, or both types of modes should be controlled and manipulated. The simulated theories considered are in 1+1 dimensions (D) but the generalization to higher dimensions, and in particular challenges involved in the case of higher-dimensional LGTs, are briefly discussed. Numerical examples are provided to supplement our proposals with concrete experimental parameters, and to demonstrate their feasibility within the current technologies. Qualitative comparisons of the simulation cost within digital and hybrid analog-digital simulations of the same theories will be provided, along with a discussion of the outlook of a hybrid approach for generalization to more complex QFTs, including non-Abelian LGTs.

\section{The hybrid analog-digital building blocks 
\label{sec:simulator}}
\noindent
The trapped-ion quantum simulator considered in this work consists of a number of ions confined in a radio-frequency Paul trap~\cite{paul1990electromagnetic}. The qubit is encoded in two stable internal levels of the ion, which are separated in energy by an angular frequency $\omega_0$.\footnote{Planck's constant $\hbar$ is set to unity throughout.} The confining potential is sufficiently stronger along the transverse axes, $\mathsf{x}$ and $\mathsf{y}$, of the trap so that the ions form a linear crystal in the axial, $\mathsf{z}$, direction.\footnote{$\mathsf{x}$, $\mathsf{y}$ and $\mathsf{z}$ should not be confused with the $x,y$ and $z$ indices to be introduced on the quasi-spins of the qubit. While the former (in san-serif font) correspond to spatial physical coordinates of the trap, hence labeling the Cartesian components of the wave-vector of the laser beams, the latter (in italic font) correspond to the Bloch-sphere axes in the qubit Hilbert space.} With appropriate anharmonic axial confinement forces, the ions can be arranged in an equally-spaced configuration for individual laser-beam addressing~\cite{lin2009large,pagano2018cryogenic}. The typical spacing between adjacent ions is a few micrometers in present-day trapped-ion simulators.

Given the long-range Coulomb force among the ions and the common trapping potential applied, the motion of the ions can be described in terms of a set of collective normal modes. The displacement of the ions around an equilibrium position can then be expressed in terms of phononic degrees of freedom, whose excitation energies are quantized in units of the normal-mode frequencies. While such a normal-mode picture is sufficient for simulating the dynamics of certain fermion-boson field theories, switching to a local-mode picture is required once the bosonic field exhibits non-trivial dynamics, as is the case in the gauge-theory example considered. Local modes can be addressed by making the trapping potential tighter along the transverse directions, or conversely by relaxing the axial confinement to increase the spacing among the ions, such that the displacement of any given ion from its equilibrium position is much smaller than the distance among the adjacent ions~\cite{porras2004bose}. The local modes can be addressed as long as the laser excitation on individual ions happens on timescales faster than the separation of the corresponding normal modes, which defines the timescale for phonon hopping.\footnote{One may also consider a micro-trap where the global trapping potential is replaced by local potentials applied at the position of each ion~\cite{cirac2000scalable}.
For applications studied in this work, a Paul trap with a tight transverse potential suffices to produce the desired dynamics, and this additional possibility will not be considered.} In such a regime, the phonon hopping probability among the ion sites is small and the quanta of motion are localized. The interactions associated with each of these scenarios, along with the expanded gate-set for the hybrid analog-digital computation of this work, will be introduced in the following.

\subsection{Phonons as excitations of normal modes of motion 
\label{sec:normal}}
The free Hamiltonian of the trapped-ion system in the absence of laser-ion interactions is
\begin{align}
&H_{\text{ion}}=\frac{\omega_0}{2}\sum_{j=1}^N \sigma_j^z+\sum_{\mathsf{m}=1}^{3N} \omega_\mathsf{m} (a_\mathsf{m}^\dagger a_\mathsf{m}+\frac{1}{2}).
\label{eq:Hfree}
\end{align}
Here, $\sigma$ is a Pauli operator acting on the space of the ions' quasi-spin, i.e., the qubit, with the standard commutation relations.  $a_\mathsf{m}^\dagger$ ($a_\mathsf{m}$) are the bosonic creation (annihilation) operators for the normal modes of motion with the associated frequencies $\omega_\mathsf{m}$ and the commutation relations $[a_\mathsf{m},a_{\mathsf{m}'}]=[a_\mathsf{m}^\dagger,a_{\mathsf{m}'}^\dagger]=0$ and $[a_\mathsf{m},a_\mathsf{m'}^\dagger]=\delta_{\mathsf{m},\mathsf{m}'}$. We label the transverse modes of motion along the $\mathsf{x}
$-axis of the trap by indices $\mathsf{m}=1,2,\cdots,N$, those along the $\mathsf{y}$ axis of the trap by indices $\mathsf{m}=N+1,N+2,\cdots,2N$, and the axial modes, i.e., those along the $\mathsf{z}$ axis of the trap, by indices $\mathsf{m}=2N+1,2N+2,\cdots,3N$. With the introduction of two counter-propagating laser beams with wave-vector difference $\Delta \bm{k}_j$, frequency difference (beatnote) $\Delta \omega_j \equiv \omega_j^L$, and phase difference $\Delta \phi_j \equiv \phi_j$, detuned from an excited internal level of the ion, the two-photon Raman transition among the two states of the qubit can be induced with a Rabi frequency $\Omega_j$ at the location of ion $j$. The beams are assumed to address the ions individually, hence the subscript $j$. In an interaction picture that rotates with the free Hamiltonian in Eq.~(\ref{eq:Hfree}), the interacting ion-laser Hamiltonian can be written as~\cite{schneider2012experimental}
\begin{align}
&H_{\text{ion-laser}}'=\sum_{j=1}^N\frac{\Omega_j}{2} e^{i\sum_{\mathsf{m}=1}^{3N}\eta_{\mathsf{m},j}(a_\mathsf{m}e^{-i\omega_\mathsf{m} t}+a_\mathsf{m}^\dagger e^{i\omega_\mathsf{m} t})}\times
\nonumber\\
&\hspace{3.8 cm}e^{-i(\omega_j^L-\omega_0)t+i\phi_j} \sigma^+_j+\text{h.c}.,
\label{eq:Hionlaser}
\end{align}
where the condition $|\omega_j^L-\omega_0| \ll \omega_0$ is assumed. The prime on $H$ is to denote that this Hamiltonian is in the interaction picture. Later on, we need to adopt a different interaction picture rotating with shifted frequencies and so it is important to bear in mind the origin of Eq.~(\ref{eq:Hionlaser}). Multiple beatnote frequencies, amplitudes, and phases can be applied simultaneously, hence a sum over laser parameters can be introduced, a possibility that we take advantage of later. The Lamb-Dicke parameters $\eta_{\mathsf{m},j}$ are defined as $\eta_{\mathsf{m},j}=\sqrt{\tfrac{(\Delta \bm{k})^2}{2M_{\rm ion}\omega_\mathsf{m}}}b_{\mathsf{m},j}$, where $b_{\mathsf{m},j}$ are the (normalized) normal-mode eigenvector components between ion
$j$ and mode $\mathsf{m}$, and $M_{\rm ion}$ denotes the mass of the ion.
\begin{figure*}[t!]
\includegraphics[scale=0.65]{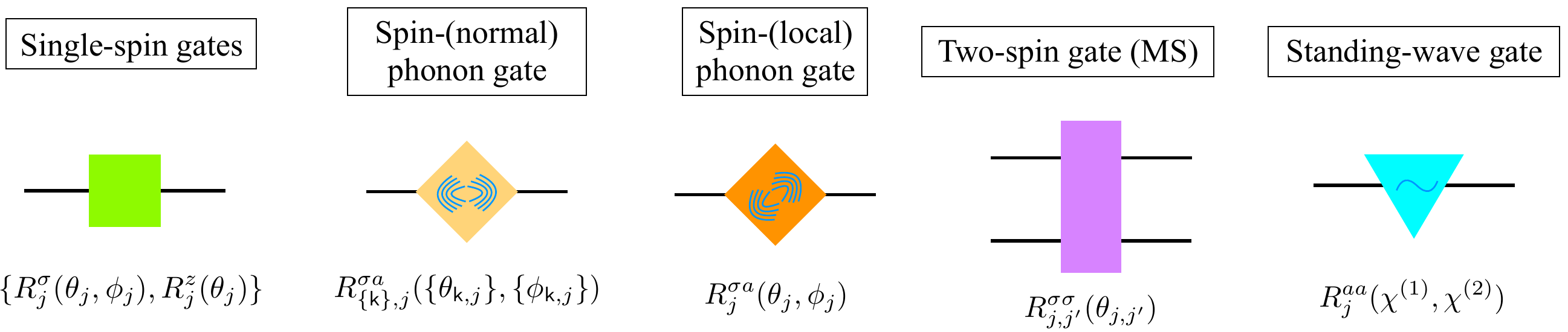}
\caption[.]{The building blocks of the analog-digital circuits in a trapped-ion quantum simulator. The gates from left to right are defined in Eqs.~(\ref{eq:Rsigma}), (\ref{eq:Rsigmaz}), (\ref{eq:Rsigmaa}),  (\ref{eq:Rsigmaalocal}), (\ref{eq:Rsigmasigma}), and (\ref{eq:Raa}), respectively.
}
\label{fig:gates}
\end{figure*}
In the Lamb-Dicke regime in which $\eta_{\mathsf{m},j}\braket{(a_\mathsf{m}+a_\mathsf{m}^\dagger)^2}^{1/2}\ll1$, transitions in the space of coupled spin-phonon system take a simpler form, and can be realized through quantum gates. $\Delta \bm{k}_j$ is assumed to be the same for lasers addressing each ion $j$, i.e., $\Delta \bm{k}_j \equiv \Delta \bm{k}$.

The carrier transition corresponds to $\left . H'_{\text{ion-laser}}\right|_{\mathcal{O}(\eta^0)}$ and is obtained by setting $\omega_j^L-\omega_0=0$. With this setting, the Hamiltonian corresponding to ion $j$ becomes
\begin{equation}
H^{\sigma}_{j}(\phi_j)=\frac{\Omega_j}{4} \bigg[(e^{i\phi_j}+e^{-i\phi_j})\sigma_j^x+(e^{i\phi_j}-e^{-i\phi_j})i\sigma_j^y \bigg].
\label{eq:Hsigma}
\end{equation}
The superscript on $H^{\sigma}_{j}$ denotes that this Hamiltonian only acts on the qubit space with the noted $\sigma$ operators. The single-spin rotations of arbitrary angle $\theta_j$, to be denoted as $R^{\sigma}_{j}(\theta_j,\phi_j)$, can be obtained by applying $H^{\sigma}_{j}(\phi_j)$ for time $\tau_{\rm gate}^{\sigma}$, to be deduced from the relation
\begin{align}
&R^{\sigma}_{j}(\theta_j,\phi_j) \equiv e^{-i\theta_j (\cos \phi_j \sigma_j^x-\sin \phi_j \sigma_j^y)}=e^{-iH_{j}^{\sigma}(\phi_j) \tau_{\rm gate}^{\sigma}}.
\label{eq:Rsigma}
\end{align}
For example, rotations around the $x$ and $y$ axes of the Bloch sphere of the qubit $j$ can be realized as
\begin{align}
&R^{\sigma}_{j}(\theta_j,0) \equiv e^{-i\theta_j \sigma_j^x},
\\
%
%
&R^{\sigma}_{j}(\theta_j,\frac{3\pi}{2}) \equiv e^{-i\theta_j \sigma_j^y},
\end{align}
respectively. Rotations along the $z$ axis of the Bloch sphere, $R^z_{j}(\theta_j)$, defined as 
\begin{align}
&R^z_{j}(\theta_j) \equiv e^{-i\theta_j \sigma_j^z}
\label{eq:Rsigmaz}
\end{align}
can be implemented without the need for a quantum operation using a classical phase shift on the controller for addressing beam $j$. A useful gate in the circuits constructed in the following sections is the phase gate $S_j$, which can be implemented via $S_j=R^z_j(\pi/4)$.

The blue and red sideband transitions correspond to $\left . H'_{\text{ion-laser}}\right|_{\mathcal{O}(\eta)}$ and are obtained by setting $\omega_j^L-\omega_0=\omega_\mathsf{k}$ and $-\omega_\mathsf{k}$, respectively. This setting leads to a coupled spin-phonon Hamiltonian. In order to achieve a Hamiltonian proportional to e.g., $\sigma^y_j$, beatnotes associated to blue and red sidebands can be applied simultaneously with phases that add up to zero~\cite{monroe2019programmable}.\footnote{To achieve a Hamiltonian of the same type that is instead proportional to $\sigma^x_j$, the red and blue sideband laser phases can be tuned to add up to $\pi$. Finally, phonon-ion Hamiltonian proportional to $\sigma^z_j$ can be obtained in a circuit by applying appropriate single-qubit rotations, as is demonstrated in later sections.} With this setting, the Hamiltonian corresponding to ion $j$ becomes
\begin{align}
&H^{\sigma a}_{\mathsf{k},j}(\phi_{\mathsf{k},j})=-\frac{\eta_{\mathsf{k},j}\Omega_j}{2} (e^{i\phi_{\mathsf{k},j}}a_\mathsf{k}+e^{-i\phi_{\mathsf{k},j}}a_\mathsf{k}^\dagger) \, \sigma_j^y,
\label{eq:Hsigmaa-single}
\end{align}
where $\phi_{\mathsf{k},j}=\frac{1}{2}(\phi^r_{\mathsf{k},j}-\phi^b_{\mathsf{k},j})=\phi^r_{\mathsf{k},j}$, with $\phi^{r(b)}_{\mathsf{k},j}$ denoting the red (blue) sideband laser phase. In order to ensure only one set of the phonon modes, $a_\mathsf{k}$, couples to the qubit in this setup, the Raman beams can be set such that the wave-vector difference $\Delta \bm{k}$ is parallel to one of the principal axes of the trap. For example, by setting $\Delta \bm{k} = \Delta k\,\hat{\bm{\mathsf{x}}}$, the beams only couple to transverse normal modes along the $\mathsf{x}$ axis. If this condition is not met and more than one set of modes are coupled, the different frequencies of the modes can be addressed with lower power to  avoid simultaneous coupling to other sets of modes. By applying multiple beatnote frequencies associated with red and blue sideband transitions, corresponding to the set $\{ \omega^L_j\}-\omega_0=\pm \{\omega_\mathsf{k}\}$ for $k \in \{1,\cdots,N\}$, and given the flexibility to set the amplitude and phase associated with each beatnote independently, the Hamiltonian
\begin{align}
&H^{\sigma a}_{\{\mathsf{k}\},j}(\{\phi_{\mathsf{k},j}\})=-\sum_\mathsf{k} \frac{\eta_{\mathsf{k},j}\Omega_{\mathsf{k},j}}{2} (e^{i\phi_{\mathsf{k},j}}a_\mathsf{k}+e^{-i\phi_{\mathsf{k},j}}a_\mathsf{k}^\dagger) \, \sigma_j^y,
\label{eq:Hsigmaa-multiple}
\end{align}
can be implemented directly. Note that $\Omega_j$ is promoted to $\Omega_{\mathsf{k},j}$ to reflect the freedom in the choice of mode-dependent Rabi frequencies. The spin-phonon rotations of an arbitrary set of angles $\{\theta_{\mathsf{k},j}\}$, to be denoted as $R^{\sigma a}_{\{\mathsf{k}\},j}(\{\theta_{\mathsf{k},j}\},\{\phi_{\mathsf{k},j}\})$, can be obtained by applying $H^{\sigma a}_{\{\mathsf{k}\},j}(\{\phi_{\mathsf{k},j}\})$ for time $\tau_{\rm gate}^{\sigma a}$, to be deduced from the relation
\begin{align}
&R^{\sigma a}_{\{\mathsf{k}\},j}(\{\theta_{\mathsf{k},j}\},\{\phi_{\mathsf{k},j}\}) \equiv e^{-i\sum_\mathsf{k}\theta_{\mathsf{k},j} (e^{i\phi_{\mathsf{k},j}}a_\mathsf{k}+e^{-i\phi_{\mathsf{k},j}}a_\mathsf{k}^\dagger)\sigma_j^y}
\nonumber\\
&\hspace{2.15 cm}=e^{-iH_{\{\mathsf{k}\},j}^{\sigma a}(\{\phi_{\mathsf{k},j}\}) \tau_{\rm gate}^{\sigma a}}.
\label{eq:Rsigmaa}
\end{align}
In particular with a single-mode addressing, opportunely setting laser-beam phases gives rise to couplings to different phonon-operator combinations proportional to $a_{\mathsf{k}}+a_{\mathsf{k}}^\dagger$ and $a_{\mathsf{k}}-a_{\mathsf{k}}^\dagger$:
\begin{align}
&R^{\sigma a}_{\mathsf{k},j}(\theta_{\mathsf{k},j},0) \equiv e^{-i\theta_{\mathsf{k},j} (a_{\mathsf{k}}+a_{\mathsf{k}}^\dagger)\sigma_j^y},
\\
&R^{\sigma a}_{\mathsf{k},j}(\theta_{\mathsf{k},j},\frac{\pi}{2}) \equiv e^{\theta_{\mathsf{k},j} (a_{\mathsf{k}}-a_{\mathsf{k}}^\dagger)\sigma_j^y}.
\end{align}

Implementing simultaneous blue and red sideband transitions detuned from the normal mode frequencies will not generate dynamical phonons as long as the lasers are applied for certain time durations. The spin-phonon couplings in this process then effectively induce a spin-spin interaction, which leads to the well-known M\o lmer-S\o rensen (MS) gate~\cite{molmer1999multiparticle, milburn2000ion, solano1999deterministic}.\footnote{Or a geometrical phase gate in case of spin interactions proportional to $\sigma^z \otimes \sigma^z$~\cite{cirac1995quantum}.} The corresponding Hamiltonian between ions $j$ and $j'$ is
\begin{align}
&H^{\sigma \sigma}_{j,j'}=\Omega_j \Omega_{j'} \sum_\mathsf{m} \eta_{\mathsf{m},j} \eta_{\mathsf{m},j'}\frac{\omega_\mathsf{m}}{(\omega_j^L-\omega_0)^2-\omega_\mathsf{m}^2}\sigma_j^x\sigma_{j'}^x,
\label{eq:Hsigmasigma}
\end{align}
Spin-spin rotations $R_{j,j'}^{\sigma\sigma}(\theta_{j,j'})$ of arbitrary angle $\theta_{j,j'}$ can be obtained by applying $H^{\sigma \sigma}_{j,j'}$ for time $\tau_{\rm gate}^{\sigma\sigma}$ given by the relation
\begin{align}
&R_{j,j'}^{\sigma\sigma}(\theta_{j,j'}) \equiv e^{-i \theta_{j,j'} \sigma^x_j \sigma^x_{j'}}=e^{-iH^{\sigma \sigma}_{j,j'}\tau_{\rm gate}^{\sigma\sigma}}.
\label{eq:Rsigmasigma}
\end{align}
In reality, designing high-fidelity two-qubit entangling gates of this type requires complex pulse shaping techniques to close all the mode trajectories in phase space, which is necessary to eliminate any residual spin-phonon coupling, and to leave the system in the same phonon state before and after the gate operation, see e.g., Refs.~\cite{Zhu06PRL, Roos_2008, Green2015, Leung2018, Blumel21}.

\subsection{Phonons as excitations of local modes of motion 
\label{sec:local}}
In a linear Paul trap, the Coulomb energy is comparable to the potential energy associated with the axial motion of the ions. Namely, the parameter $\beta^\mathsf{z} \equiv e^2/(d_0^3M_\text{ion}{\omega^\mathsf{z}}^2) \gtrsim 1$, where $\omega^\mathsf{z}$ is the trap frequency along the axial axis of the trap, $d_0$ is the average spacing among the ions, and $e$ is the electric charge. As a result, the phonons are shared across the chain as excitations of collective normal modes. The same parameter is typically much smaller for the transverse motions of the ions. In particular, if the trap is made extremely tight along the transverse axes of the trap such that $\beta^{\mathsf{x}(\mathsf{y})} \ll 1$, the transverse phonons describe the oscillation of a single ion, with a relatively small coupling strength for hopping to the neighboring ions that is $\propto e^2/(d_0^3M_\text{ion}\omega^{\mathsf{x}(\mathsf{y})})=\beta^{\mathsf{x}(\mathsf{y})} \omega^{\mathsf{x}(\mathsf{y})}$~\cite{porras2004bose}. For the quantum simulation of QFTs that exhibit non-trivial boson-field dynamics, it is necessary to adopt a local-mode picture. Assuming that the transverse directions of the trap support local phonon modes, the free Hamiltonian of the ion chain can be written as
\begin{align}
&H_{\text{ion}}=\frac{\omega_0}{2}\sum_{j=1}^N \sigma_j^z+\sum_{j=1}^{N} \omega^\mathsf{x}_j ({a_j^\mathsf{x}}^\dagger a^\mathsf{x}_j+\frac{1}{2})\,+
\nonumber\\
&\hspace{0.4 cm}
\sum_{j=N+1}^{2N} \omega^\mathsf{y}_j ({a_j^\mathsf{y}}^\dagger a^\mathsf{y}_j+\frac{1}{2})+\sum_{\mathsf{m}=2N+1}^{3N} \omega_\mathsf{m} (a_\mathsf{m}^\dagger a_\mathsf{m}+\frac{1}{2}),
\label{eq:HionLocal}
\end{align}
where in the second and third terms, the subscript $j$ on the phonon operators is a local index corresponding to the quanta of motion of the $j^{\rm th}$ ion.

For the field theories considered in this work, the simulated theory does not exhibit boson hopping and so the phonon hopping in the simulator needs to be suppressed. In a typical trapped-ion system with tens of Ytterbium ions, $\omega^\mathsf{x} \sim 2\pi \times 5$ MHz and $d_0$  is of the order of a few microns. Then $\beta^{\mathsf{x}} \sim 10^{-3}-10^{-4}$ and $\beta^\mathsf{x} \omega^\mathsf{x} \sim 2\pi \times 1$ kHz, and hence the dynamics associated with the phonon hopping can be neglected compared with spin-phonon (with typical strength $\sim 100$ kHz) and spin-spin dynamics (with typical strength $\sim 10$ kHz). If further suppression of the phonon hopping is desired to improve the accuracy of the simulated model, one can either impose tighter trapping potential along the transverse directions or spread out the ions further by reducing the axial potential. Additionally, the nearest-neighbor hopping can be actively blocked using techniques demonstrated in Ref.~\cite{debnath2018observation,Ohira21} or by applying local optical tweezers to vary the local confinement~\cite{olsacher2020scalable, shen2020scalable, espinoza2021engineering, teoh2021manipulating}. Another option, which does not require additional blockade ions, is the use of a mixed-species ion chain with a large mass ratio, in which the modes for the different species separate by mode participation~\cite{sosnova2021character}. In an alternating arrangement, phonon hopping will be suppressed by the local-mode frequency difference between neighboring ions of different species and the increased distance between the ions of the same species. Finally, we note that the local-mode frequencies can be expressed as the common trap frequency plus additional local corrections. These corrections scale similar to the hopping strength and are hence suppressed by at least a factor of $10^{-3}$ compared with the trap frequency in a typical trap described above. As a result, $\omega_j^\mathsf{x}$ in Eq.~(\ref{eq:HionLocal}) can be replaced with $\omega^\mathsf{x}$ to a first approximation.

The spin-phonon Hamiltonians in Eq.~(\ref{eq:Hsigmaa-single}) can be obtained similarly by expanding the ion-laser interaction Hamiltonian in Eq.~(\ref{eq:Hionlaser}) using the local modes of motion. Then both the spin $\sigma_j$ and the phonon operators ($a^\mathsf{x}_j$ and ${a^\mathsf{x}_j}^\dagger$) are representing quantum operations at location $j$ along the chain, with the associated gate operation identified as
\begin{align}
&R^{\sigma a}_j(\theta_j,\phi_j) = R^{\sigma a}_{\mathsf{k},j}(\theta_{\mathsf{k},j},\phi_{\mathsf{k},j})\big |_{a_\mathsf{k} \to a^\mathsf{x}_j,\eta_{\mathsf{k},j} \to \eta_j,\theta_{\mathsf{k},j} \to \theta_j,\phi_{\mathsf{k},j} \to \phi_j},
\label{eq:Rsigmaalocal}
\end{align}
where the single-mode addressing limit of Eq.~(\ref{eq:Rsigmaa}) is considered. On the other hand, in the limit of a tight transverse direction where the hopping of the phonons along transverse axes of the trap is suppressed, the spin-spin gate in Eq.~(\ref{eq:Rsigmasigma}) can be still derived with high fidelity using the normal modes of motion along the axial direction so to have sufficiently strong all-to-all spin-spin couplings, given that $\beta^z\gg\beta^{x(y)}$. As will be shown, both types of gates, one based on local modes of motions and the other based on normal modes of motion, can be employed in the simulation, exhibiting another advantage of a hybrid analog-digital setting.

For the simulation of the lattice Schwinger model, the bosons need to interact locally. Such an interaction will need to be mapped to a local phonon-phonon interaction. To engineer this new type of interaction in the trapped-ion simulator, as proposed in Ref.~\cite{porras2004bose}, the ions can be placed near the minima (or maxima) of a standing wave, which induces an AC Stark shift corresponding to the Hamiltonian
\begin{align}
&H^{\text{s.w.}}_j=F \cos^2(\widetilde{k}\mathsf{x}_j)=F\cos^2\big[\widetilde{\eta} (a^\mathsf{x}_j+{a^\mathsf{x}_j}^\dagger)\big]
\nonumber\\
&\hspace{0.85 cm}=2F(-\widetilde{\eta}^2+\widetilde{\eta}^4)({a^\mathsf{x}_j}^\dagger a^\mathsf{x}_j)+2F\widetilde{\eta}^4({a^\mathsf{x}_j}^\dagger a^\mathsf{x}_j)^2+\cdots.
\label{eq:Hsw}
\end{align}
Here, $\widetilde{\bm{k}}=\widetilde{k} \, \hat{\bm{\mathsf{x}}}$ where $\widetilde{k}$ is the wave-vector of the standing-wave beam and $\widetilde{\eta}\equiv \widetilde{k}\mathsf{x}^{(0)}= \sqrt\frac{\widetilde{k}^2}{2M_{\rm ion}\omega^\mathsf{x}}$ is the corresponding Lamb-Dicke parameter. Note that  the standing wave is only exciting the local modes of motion along one direction based on the choice of the wave-vector. Eq.~(\ref{eq:Hsw}) does not consider higher-order terms in $\widetilde{\eta}^2$ that can be neglected in the regime $\widetilde{\eta}\braket{(a^\mathsf{x}_j+{a^\mathsf{x}}^\dagger_j)^2}^{1/2}\ll1$. It also neglects phonon non-conserving operators that rotate with at least the frequency $\omega^\mathsf{x}$, and can be adiabatically eliminated as long as $F\widetilde{\eta}^2/\omega^\mathsf{x} \ll 1$~\cite{porras2004bose}.

Analogous to the previous gates, phonon-phonon rotations of arbitrary angles can be obtained by applying $H^{\text{s.w.}}_j$ for time $\tau_{\rm gate}^{aa}$, to be deduced from the relation
\begin{align}
&R^{aa}_{j}(\chi^{(1)},\chi^{(2)}) \equiv e^{-i \big[\chi^{(1)}({a^\mathsf{x}_j}^\dagger a^\mathsf{x}_j)+\chi^{(2)}({a^\mathsf{x}_j}^\dagger a^\mathsf{x}_j)^2\big]}
\nonumber\\
&\hspace{2.2 cm}=e^{-iH_{j}^{\rm s.w.} \tau_{\rm gate}^{aa}}.
\label{eq:Raa}
\end{align}
An important point to notice regarding this gate is that according to Eq.~(\ref{eq:Hsw}), the relative size of the coefficients of the ${a^\mathsf{x}_j}^\dagger a^\mathsf{x}_j$ and $({a^\mathsf{x}_j}^\dagger a^\mathsf{x}_j)^2$ terms, $\chi^{(1)}/\chi^{(2)}$, is fixed to $(-1+\widetilde{\eta}^2)/\widetilde{\eta}^2$, while the overall strength of these terms can be tuned arbitrarily by changing $F$ via the standing-wave intensity, as long as the condition $F\widetilde{\eta}^2/\omega^\mathsf{x} \ll 1$ is not violated. The ratio constraint may seem too limiting for simulating an arbitrary Hamiltonian containing both phonon terms. However, as shown in Sec.~\ref{sec:U(1)}, the desired ratio of the coefficients in the simulation of the lattice Schwinger model can be engineered  by appropriately choosing the laser frequencies, which amounts to choosing a suitable interaction picture. Last but not least, if the simulated theory requires site-dependent coefficients for the phonon-phonon couplings, individual standing-wave beams could replace the global beam in Eq.~(\ref{eq:Hsw}), at the expense of added experimental complexity.

\section{A Yukawa theory: scalar fields coupled to (staggered) fermions
\label{sec:Yukawa}}
\noindent
A scalar field theory coupled to fermions describes the coupling of the Higgs boson to fermions of the Standard Model through Yukawa interactions, and is responsible for dynamical mass generation and fermion mixing in nature. Non-perturbative studies of the Yukawa theory using lattice-regularized Euclidean field-theory simulations have drawn considerable interest~\cite{Fodor:1994sj, Gerhold:2007yb, Gerhold:2009ub, Bulava:2012rb, Chu:2017vmc, Jansen:2014lja, Brambilla:2014jmp, Akerlund:2016xek} as they reveal important connections to the cutoff scale of the Standard Model and questions regarding quantum triviality~\cite{Callaway:1988ya}. In the context of cosmology and the early universe, non-perturbative simulations are required for studies of non-equilibrium and real-time dynamics of this theory and its beyond-the-Standard-Model cousins~\cite{Boyanovsky:2001va}. A scalar field theory coupled to (non-relativistic) fermions is also the effective field theory description of the Yukawa interactions among the nucleons and pions, and enters the quantum many-body description of nuclei~\cite{Weinberg:1991um, Machleidt:2011zz, Kaplan:1998we}. It is therefore highly relevant to investigate the prospects for quantum simulation of such theories, including the suitability of the hybrid analog-digital approach. In the following, we focus on a simple case: a scalar field theory coupled to one flavor of staggered fermions in 1+1 D and without the possibility of self-interactions among the scalar fields. Later, we comment on the applicability of this proposal for self-interacting scalar fields and the higher-dimensional case, and will point out the requisite extensions.
\begin{figure*}[t!]
\includegraphics[scale=0.6]{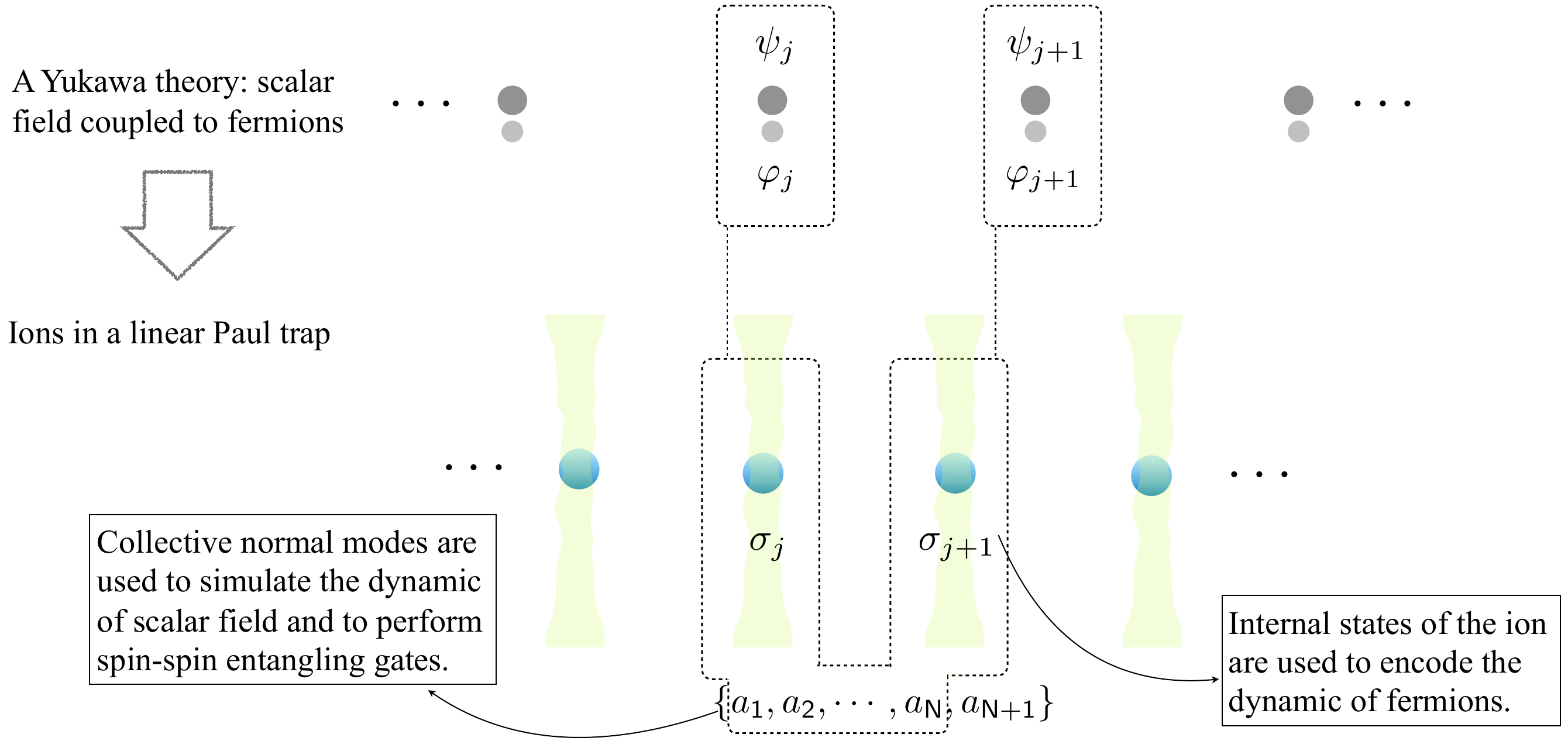}
\caption[.]{The degrees of freedom of the lattice-regularized scalar field theory coupled to staggered fermions in 1+1 D (top row) are mapped to those in a linear trapped-ion quantum simulator (bottom row) with individual laser-beam addressing. This scheme involves only normal modes of motion.
}
\label{fig:scalar-mapping}
\end{figure*}
\subsection{The Yukawa model}
Consider a one-dimensional spatial lattice with $N$ sites, lattice spacing $b$, and with periodic boundary conditions (PBCs) imposed on the fields.\footnote{Other boundary conditions can be studied with minimal adjustments.} The Hamiltonian of the lattice-regularized Yukawa theory to be simulated with a trapped-ion quantum simulator consists of 
\begin{align}
&H_{\text{Yukawa}}=H_{\text{Yukawa}}^{(I)}+H_{\text{Yukawa}}^{(II)}+H_{\text{Yukawa}}^{(III)},
\label{eq:HY}
\end{align}
where the purely-fermionic Hamiltonian
\begin{align}
&H_{\text{Yukawa}}^{(I)}=\sum_{j=1}^N\bigg[\frac{i}{2b}(\psi^\dagger_j \psi_{j+1}-\psi^\dagger_{j+1} \psi_j)+m_{\psi}(-1)^j \psi^\dagger_j \psi_j \bigg]
\label{eq:HYI}
\end{align}
describes the hopping term and the mass term of one flavor of staggered fermions with mass $m_{\psi}$. Note that PBCs impose the identification $\psi_{N+1} \equiv \psi_1$. The free scalar-field Hamiltonian is
\begin{align}
&H_{\text{Yukawa}}^{(II)}=b\sum_{j=1}^N\bigg[\frac{\Pi_j^2}{2}+\frac{(\nabla\varphi_j)^2}{2}+\frac{m^2_{\varphi}}{2}\varphi_j^2\bigg],
\label{eq:HYII1}
\end{align}
where $\Pi_j$ is the conjugate momentum corresponding to the scalar field $\varphi_j$, i.e. $[\varphi_j,\varphi_{j'}]=[\Pi_j,\Pi_{j'}]=0$ and $[\varphi_j,\Pi_{j'}]=i(Nb)^{-1}\delta_{j,j'}$. These fields can be quantized in the standard way to obtain a representation in terms of the (bosonic) harmonic-oscillator creation ($d_k^\dagger$) and annihilation ($d_k$) operators,
\begin{align}
&\varphi_j=\frac{1}{\sqrt{Nb}}\sum_{k=-N/2}^{N/2-1} \frac{1}{\sqrt{2\varepsilon_k}}(d_k^\dagger e^{-i2\pi kj/N}+d_k e^{i2\pi kj/N}),
\label{eq:phi}
\\
%
%
%
&\Pi_j=\frac{i}{\sqrt{Nb}}\sum_{k=-N/2}^{N/2-1} \sqrt{\frac{\varepsilon_k}{2}}(d_k^\dagger e^{-i2\pi kj/N}-d_k e^{i2\pi kj/N}),
\label{eq:pi}
\end{align}
where $k$ labels the corresponding momentum mode $p_k \equiv 2\pi k/(Nb)$, and  $\varepsilon_k=\sqrt{(\frac{2\pi k}{Nb})^2+m_\varphi^2}$ is the corresponding energy, with $m_\varphi$ being the bare mass of the scalar field. Inputting Eqs.~(\ref{eq:phi}) and (\ref{eq:pi}) in Eq.~(\ref{eq:HYII1}), and using the commutation relations of the bosonic operators: $[d_k,d_{k'}]=[d_k^\dagger,d_{k'}^\dagger]=0$ and $[d_k,d_{k'}^\dagger]= \delta_{k,k'}$, one arrives at the well-known Hamiltonian
\begin{align}
&H_{\text{Yukawa}}^{(II)}=\sum_{k=-N/2}^{N/2-1}\varepsilon_k \, (d^\dagger_k d_k+\frac{1}{2}),
\label{eq:HYII2}
\end{align}
describing the energy of $N$ uncoupled quantum harmonic oscillators. Finally, the interacting fermion scalar-field Hamiltonian is
\begin{align}
&H_{\text{Yukawa}}^{(III)}=gb\sum_{j=1}^N \psi^\dagger_j \varphi_j \psi_j,
\label{eq:HYIII}
\end{align}
where field $\varphi$ must be realized as the collection of quantum harmonic oscillators through Eq.~(\ref{eq:phi}).

\subsection{The mapping to an analog-digital circuit
\label{sec:Yukawa-mapping}
}
To map the Hamiltonian in Eq.~(\ref{eq:HY}) to the building blocks of the analog-digital trapped-ion simulators introduced in Sec.~\ref{sec:simulator}, one may first transform the staggered fermionic fields straightforwardly to the spin operators through a Jordan-Wigner transformation: $\psi_i = \prod_{l<j}(i\sigma_l^z)\sigma_j^-$ and $\psi_i^\dagger = \prod_{l<j}(-i\sigma_l^z)\sigma_j^+$ with $\sigma_j^\pm=\frac{1}{2}(\sigma_j^x \pm i \sigma_j^y)$. Additionally, the harmonic oscillators can be mapped to the phonons associated with the normal modes of the motion (in either the axial or one of the transverse directions depending on the convenience of the experimental setting). Explicitly, with the labeling rule defined after Eq.~(\ref{eq:Hfree}), the mapping to the transverse normal modes along the $\mathsf{x}$ direction reads: $d_k := a_{k+N/2+1}$ and $d^\dagger_k := a^\dagger_{k+N/2+1}$ for $k=-N/2,-N/2+1,,\cdots,N/2-1$. To keep the occupation of these modes unaffected while implementing spin-spin entangling gates, the other set of transverse modes or the axial modes can be addressed to implement the MS gates. With the degrees of freedom in the simulated theory being mapped to qubit and phonon degrees of freedom of the trapped-ion simulator, what is left to identify are the gate operations that implement $e^{-iH_\text{Yukawa}t}$. Given the digital setting of this proposal, the time-evolution operator can be digitized using the lowest-order Trotter-Suzuki expansion~\cite{suzuki1991general}. Higher-order expansions~\cite{wiebe2010higher, childs2019theory}, as well other state-of-the-art simulation algorithms~\cite{berry2007efficient, berry2009black, childs2012hamiltonian, berry2014exponential, berry2015hamiltonian, berry2015simulating, low2018hamiltonian, low2019hamiltonian, kalev2020simulating}, can be similarly investigated within the current approach, but will not be discussed further in this work. 

To map the Hamiltonian in Eq.~(\ref{eq:HY}) to the Hamiltonians associated with the gates introduced in the previous section, a re-arrangement of the terms is performed such that $H_{\text{Yukawa}}$ is now broken down to
\begin{align}
&H_{\text{Yukawa}}^{(I)'}=\frac{1}{4b}\sum_{j=1}^N \sigma_j^x\sigma_{j+1}^x,
\label{eq:HYIprime}
\\
&H_{\text{Yukawa}}^{(II)'}=\frac{1}{4b}\sum_{j=1}^N \sigma_j^y\sigma_{j+1}^y,
\label{eq:HYIIprime}
\\
%
%
&H_{\text{Yukawa}}^{(III)'}=\frac{m_\psi}{2} \sum_{j=1}^N (-1)^j\sigma_j^z+{\rm const.},
\label{eq:HYIIIprime}
\end{align}
\begin{align}
&H_{\text{Yukawa}}^{(IV)'}=\sqrt{\frac{g^2b}{8N}}\sum_{j=1}^N (\mathbb{I}_j+\sigma_j^z)\sum_{\mathsf{m}=1}^{N} \frac{1}{\sqrt{\varepsilon_\mathsf{m}}}\,\times
\nonumber
\\
%
%
&\hspace{1.95 cm} (a_\mathsf{m}^\dagger e^{-i\frac{2\pi j}{N}(\mathsf{m}-\frac{N}{2}-1)}+a_\mathsf{m} e^{i\frac{2\pi j}{N}(\mathsf{m}-\frac{N}{2}-1)})\,+
\nonumber\\
&\hspace{4.5 cm} 
\sum_{\mathsf{m}=1}^{N} \varepsilon_\mathsf{m} (a_\mathsf{m}^\dagger a_\mathsf{m}+\frac{1}{2}),
\label{eq:HYIVprime}
\end{align}
where the mapping to spin and phonon degrees of freedom has already been carried out. To generate the free scalar-field Hamiltonian in Eq.~(\ref{eq:HYII2}) for generic mode-dependent coefficients $\varepsilon_k \to \varepsilon_{\mathsf{m}=k+N/2+1}$, one can resort to a change of interaction-picture Hamiltonian~\cite{mezzacapo2012digital}. The gates obtained so far are in an interaction picture derived using the free Hamiltonian of the ion system in Eq.~(\ref{eq:Hfree}), containing the harmonic-oscillator energy term with mode-independent and experimentally-fixed coefficients. However, one can introduce the term $\sum_\mathsf{m} \widetilde{\varepsilon}_\mathsf{m} (a_\mathsf{m}^\dagger a_\mathsf{m}+\frac{1}{2})$ in the interacting Hamiltonian by appropriately choosing the rotating frame. $\widetilde{\varepsilon}_\mathsf{m}$ is the properly rescaled $\varepsilon_\mathsf{m}$ accounting for the ratio of the model time variable and the experimental gate time, see the description after Eq.~(\ref{eq:expHYIVprime}). In particular, setting the free Hamiltonian to $H_{\text{ion}}-\sum_\mathsf{m}\widetilde{\varepsilon}_\mathsf{m} (a_\mathsf{m}^\dagger a_\mathsf{m}+\frac{1}{2})$, the Hamiltonian in the interaction picture becomes $H'_\text{ion-laser}+\sum_\mathsf{m}\widetilde{\varepsilon}_\mathsf{m} (a_\mathsf{m}^\dagger a_\mathsf{m}+\frac{1}{2})$. Operationally, this implies adjusting the detuning of the red and blue sideband transitions, in other words tuning the laser frequency to $\omega_j^L-\omega_0=\pm (\omega_\mathsf{k}-\widetilde{\varepsilon}_\mathsf{k})$, which would implement the desired mode-dependent term $\widetilde{\varepsilon}_\mathsf{k} (a_\mathsf{k}^\dagger a_\mathsf{k}+\frac{1}{2})$ in the evolution. 

However, this is not the full picture yet, since the evolution of the Yukawa theory will be implemented in a digital manner. In particular, there will be multiple sideband operations (when implementing spin-phonon gates) during each step of the Trotter evolution, and these effectively implement the phonon-energy term as well. Therefore, one must ensure that this term will be induced with the correct coefficient. As will be seen below, implementing the time evolution of the Yukawa theory requires introducing an ancilla ion and amounts to a total of $N+1$ multi-mode spin-phonon gates per Trotter step. This necessitates changing the free Hamiltonian to $H_{\text{ion}}-\frac{1}{(N+1)}\sum_\mathsf{m}\widetilde{\varepsilon}_\mathsf{m} (a_\mathsf{m}^\dagger a_\mathsf{m}+\frac{1}{2})$, which leads to the interaction-picture Hamiltonian 
\begin{figure*}[t!]
\includegraphics[scale=0.64]{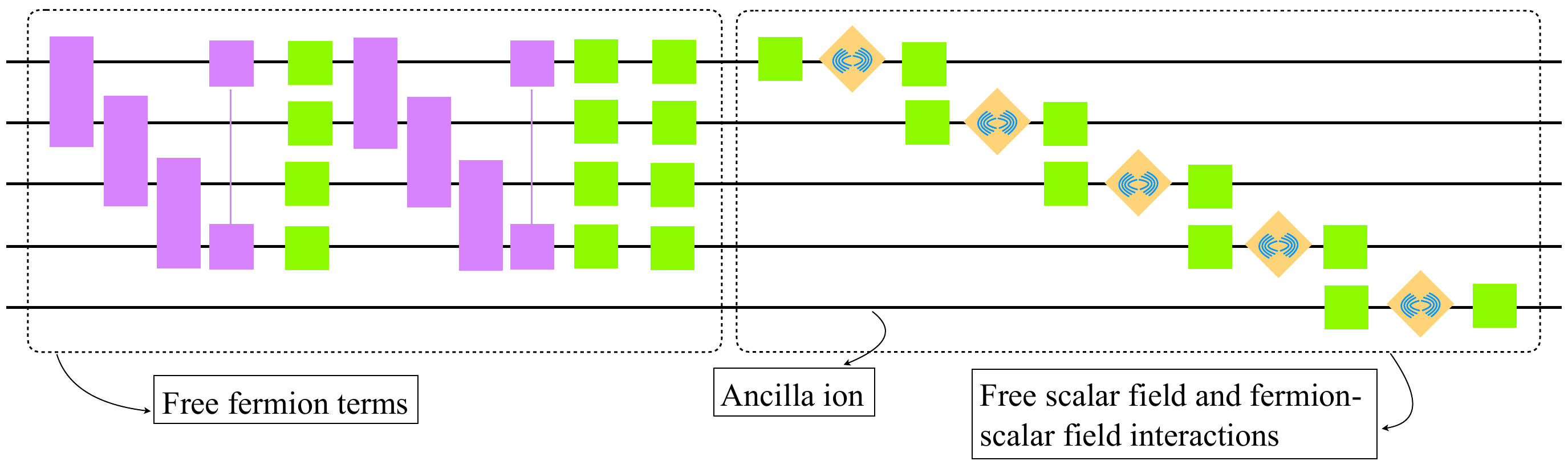}
\caption[.]{The schematic of the analog-digital quantum circuit associated with the time evolution of a four-site Yukawa theory for a single Trotter step, as expressed in Eqs.~(\ref{eq:expHYIprime})-(\ref{eq:expHYIVprime}). The gate symbols are defined in Fig.~\ref{fig:gates}.
}
\label{fig:circuit-scalar}
\end{figure*}
\begin{align}
&H_{\text{ion-laser}}''=\sum_{j=1}^N\sum_L\frac{\Omega_j^L}{2} \bigg[e^{i\sum_\mathsf{m}\eta_{\mathsf{m},j}(a_\mathsf{m}e^{-i(\omega_\mathsf{m}-\frac{\widetilde{\varepsilon}_\mathsf{m}}{(N+1)}) t}+}
\nonumber\\
&\hspace{1 cm}{}^{a_\mathsf{m}^\dagger e^{i(\omega_\mathsf{m}-\frac{\widetilde{\varepsilon}_\mathsf{m}}{(N+1)}) t})}
%
%
e^{-i(\omega_j^L-\omega_0)t+i\phi^L_j} \sigma^+_j+\text{h.c.}\bigg]+
\nonumber\\
&\hspace{2.95 cm}
\frac{1}{(N+1)}\sum_\mathsf{m=1}^N \widetilde{\varepsilon}_\mathsf{m} (a_\mathsf{m}^\dagger a_\mathsf{m}+\frac{1}{2}),
\end{align}
where $\sum_L$ is introduced to denote the possibility of simultaneous application of $N$ beatnote frequencies $\omega_j^L \equiv \omega_0\pm (\omega_\mathsf{k}-\frac{1}{N+1}\widetilde{\varepsilon}_\mathsf{k})$ for all $\mathsf{k} \in \{1,\cdots,N\}$ with their associated amplitude $\Omega_j^L \equiv \Omega_{\mathsf{k},j}$ and phase $\phi_j^L \equiv \phi_{\mathsf{k},j}$, which is taken advantage of in the application of spin-phonon gates in the following. It should be emphasized again that experimentally, the change in the interaction picture corresponds to detuning each of the spin-phonon operations. Note that the $(N+1)^{\rm th}$ normal mode is not affected by the chosen interaction picture, as its dynamics are irrelevant in simulating the scalar fields.

With these adjustments, the time evolution of the Yukawa theory for duration $\delta t$ can now be correctly obtained considering
\begin{align}
&e^{-iH_{\text{Yukawa}}' \delta t}=e^{-iH_{\text{Yukawa}}^{(I)'}\delta t} e^{-iH_{\text{Yukawa}}^{(II)'}\delta t} e^{-iH_{\text{Yukawa}}^{(III)'}\delta t}\,\times
\nonumber\\
&\hspace{3.75 cm}
e^{-iH_{\text{Yukawa}}^{(IV)'}\delta t} +\mathcal{O}\big((\delta t)^2\big),
\end{align}
and upon expressing each evolution term as 
\begin{align}
&e^{-iH_{\text{Yukawa}}^{(I)'}\delta t}=\prod_{j=1}^N \mathcal{R}_{j,j+1}^{\sigma\sigma}(\theta_{j,j+1}),
\label{eq:expHYIprime}
\\
&e^{-iH_{\text{Yukawa}}^{(II)'}\delta t}=\prod_{j=1}^N S_jS_{j+1}\mathcal{R}_{j,j+1}^{\sigma\sigma}(\theta_{j,j+1})S^\dagger_jS^\dagger_{j+1},
\label{eq:expHYIIprime}
\end{align}
with $\theta_{j,j+1}=\frac{\delta t}{4b}$,
\begin{align}
&e^{-iH_{\text{Yukawa}}^{(III)'}\delta t}=\prod_{j=1}^N R_j^z(\theta_j),
\label{eq:expHYIIIprime}
\end{align}
with $\theta_j=\frac{1}{2}m_{\psi} (-1)^j \delta t$, and
\begin{align}
&e^{-iH_{\text{Yukawa}}^{(IV)'}\delta t}=\prod_{j=1}^NR_j^\sigma(\frac{\pi}{4},0) R_{\{\mathsf{m}\},j}^{\sigma a}(\{\theta_{\mathsf{m},j}\},\{\phi_{\mathsf{m},j}\})
\nonumber\\
&\hspace{4.25 cm} \times R_j^\sigma(-\frac{\pi}{4},0) R_{N+1}^\sigma(\frac{\pi}{4},0)  \nonumber
\\
%
%
&\hspace{0.9 cm} \times R_{\{\mathsf{m}\},N+1}^{\sigma a}(\{\theta_{\mathsf{m},N+1}\},\{\phi_{\mathsf{m},j}\})R_{N+1}^\sigma(-\frac{\pi}{4},0),
\label{eq:expHYIVprime}
\end{align}
with $\theta_{\mathsf{m},j}=\theta_{\mathsf{m},N+1}=\sqrt{\frac{g^2b}{8N\varepsilon_\mathsf{m}}}\delta t$ and $\phi_{\mathsf{m},j}=\frac{2\pi j}{N}(\mathsf{m}-\frac{N}{2}-1)$, and with all values of $\mathsf{m}$ in the range $\{1,\cdots,N\}$. $R^{\sigma}_{j}(\theta_j,\phi_j)$ is defined in Eq.~(\ref{eq:Rsigma}) and its operation remains the same despite the change in the interaction-picture Hamiltonian. $R^{\sigma a}_{\{\mathsf{m}\},j}(\{\theta_{\mathsf{m},j}\},\{\phi_{\mathsf{m},j}\})$ is defined in Eq.~(\ref{eq:Rsigmaa}) but must be realized with red and blue sideband detunings $\omega_j^L-\omega_0=\pm(\omega_\mathsf{k}-\frac{1}{(N+1)}\widetilde{\varepsilon}_\mathsf{k})$ with $\widetilde{\varepsilon_\mathsf{k}}\tau_{\rm gate}^{\sigma a}=\varepsilon_\mathsf{k}\delta t$, where $\tau_{\rm gate}^{\sigma a}$ is determined from Eq.~(\ref{eq:Rsigmaa}) with the $\theta_{\mathsf{m},j}$ values specified above. It should be noted that expectation values of observables are invariant under the change in the interaction picture as long as the time-dependent states are transformed accordingly. For simple observables such as fermion and boson occupations, measurements in the original basis obtain the correct expectation values as the corresponding operators commute with the transformation~\cite{Lv:2018cxf, pedernales2015quantum}. Finally, $R_{j,j'}^{\sigma\sigma}(\theta_{j,j'})$, defined in Eq.~(\ref{eq:Rsigmasigma}), can operate using transverse normal modes other than those used in spin-phonon gates, or axial modes. Note that an ancilla qubit, labeled as $N+1$, is introduced in Eq.~(\ref{eq:expHYIVprime}) to effectively implement the interactions proportional to $\mathbb{I}_j$ in Eq.~(\ref{eq:HYIVprime})~\cite{mezzacapo2012digital}. This ancilla qubit is an extra ion prepared in the spin up state and can be used in all subsequent Trotter steps. A schematic of the circuit for a single Trotter step is shown in Fig.~\ref{fig:circuit-scalar}. We will study the benefits of this hybrid proposal for the simulation of Yukawa theory in Sec.~\ref{sec:cost-yukawa}.

The algorithm above can be generalized to a Yukawa theory in higher dimensions. In $d$ spatial dimensions, the number of sites on the lattice is $N^d$, where $N$ is the number of sites along each Cartesian direction. The number of ions required to fully encode the dynamics is $N^d$, plus a single ancilla ion as introduced in the 1+1 D case. This is because the number of normal modes of motion associated with each of the transverse (or axial) directions will also grow as $N^d$ (considering the ancilla ion, as $(N+1)^d$), which is more than sufficient to encode the $N^d$ momentum modes of the scalar field in the harmonic oscillator basis. Each ion needs to couple to all $N^d$ phonon modes, which polynomially increases the number of spin-phonon gates required to simulate the fermion scalar-field interaction term. The more significant overhead in terms of the entangling operations is caused by the encoding of the fermionic hopping term, which in the Jordan-Wigner transformation involves implementing a chain of Pauli operators, with the number of Pauli operators growing polynomially with $N$. These operations can be performed through the known decomposition into spin-spin gates, either in series or in parallel. The parallel implementation can be achieved either in one go using a global MS-operation and shelving the ions that do not participate in the coupling using individual beams~\cite{barreiro2011open}, or by more involved optical pulse-shaping methods as demonstrated in Refs.~\cite{figgatt2019parallel, lu2019global, grzesiak2020efficient}.

Another important generalization of the Yukawa theory is to incorporate self-interactions of the scalar fields, for example through a quartic interaction Hamiltonian:
\begin{align}
&H_{\text{Yukawa}}^{(IV)}=b^d\lambda \sum_j\varphi_j^4.
\label{eq:HY4}
\end{align}
This term represents non-local interactions among quantum harmonic oscillators in momentum space, requiring all-to-all phonon-phonon couplings to be engineered in the trapped-ion simulator. While inducing phonon-phonon coupling among the normal modes is possible by taking advantage of the intrinsic non-linearity of the Coulomb interaction \cite{Marquet2003} or through mediating the interaction via virtual spin degrees of freedom, such an implementation will be more challenging than the other set of gates introduced so far. The use of the local phonon modes will not be optimal either, as the phonon hopping is suppressed beyond nearest-neighbor sites, and its strength cannot be made homogeneous across the sites as required by Eq.~(\ref{eq:HY4}). Searching for efficient and feasible implementations of the Hamiltonian in Eq.~(\ref{eq:HY4}) that are more natural to a the trapped-ion simulator will be the subject of future work.

The model considered in this work is still of phenomenological interest, for example in simulating lattice effective field theory of nucleons coupled to pions. The self-interactions of pions matter only at higher orders in the chiral effective field theory of nuclear forces~\cite{Machleidt:2011zz, Kaplan:1998we} and can be neglected at low energies. Besides the trivial incorporation of multiple flavors of (non-relativistic) fermions representing spin and isospin components of a nucleon, and the (local) self-interaction of fermions representing nucleons two- and three-body contact interactions, the fermion scalar-field interacting term in Eq.~(\ref{eq:HYIII}) must be promoted to a derivative coupling in order to represent a pion-nucleon coupling. Such a term can be implemented by representing the derivative of the scalar field by a finite difference among the fields at adjacent sites. This then amounts to multiple implementations of the terms of the type in Eq.~(\ref{eq:HYIII}) that are added with appropriate signs. Further, there are three different types of pions, each requiring their own harmonic-oscillator representation (two of which being electrically charged). These can either be realized through the three sets of normal modes or by enlarging the ion chain to create more normal modes of one type for encoding of all the pionic degrees of freedom. As the building blocks of this construct are identical to the ones for the simple Yukawa model above, the explicit circuit will not be discussed further.
\begin{table}[b!]
\scalebox{1}
{
\begin{tabular}{ccccccccc}
\hline 
\multicolumn{9}{c}{Model parameters}\tabularnewline
\hline 
\hline 
 &  &  &  &  &  &  &  & \tabularnewline
$b$ & $N$ & $\Lambda$ & $g$ & $m_{\psi}$ & $m_{\varphi}$ & $\varepsilon_{\mathsf{m}}$ & $\delta t$ & $t$\tabularnewline
 &  &  &  &  &  &  &  & \tabularnewline
\hline 
 &  &  &  &  &  &  &  & \tabularnewline
1 & 2 & 8 & $5\sqrt{2}$ & 1 & 1 & $\left\{ 3.297,1\right\} $ & 0.25 & 5\tabularnewline
 &  &  &  &  &  &  &  & \tabularnewline
\hline 
 &  &  &  &  &  &  &  & \tabularnewline
1 & 4 & 1 & 5 & 1 & 1 & $\left\{ 3.297,1.862,1,1.862\right\} $ & 0.125 & 2.5\tabularnewline
 &  &  &  &  &  &  &  & \tabularnewline
\hline 
\end{tabular}
}
\caption{The parameters of the Yukawa theory considered in the two examples of this section, each corresponding to different $N$ and $\Lambda$ values as noted. The corresponding trap and gate parameters are tabulated in Appendix~\ref{app:yukawa}. $t$ is the total evolution time and $\delta t$ is the duration of the Trotter-evolution segment.}
\label{tab:yukawa-param}
\end{table}
\subsection{An example with realistic parameters
\label{sec:yukawa-example}}
To demonstrate the viability of near-term experimental implementations of the dynamics in this model with a trapped-ion simulator, one can investigate the range of the gate parameters required to observe interesting phenomena in this model. While only small instances of the problem can be studied classically, as is evident from the examples considered, the quantum simulator with tens of ions can still operate in the same gate-parameter range and already push the limits of classical capabilities, particularly given the native implementation of boson dynamics in the simulator. Later on in Sec.~\ref{sec:cost}, we demonstrate the advantage of an analog-digital approach by making a qualitative cost comparison with a fully-digital implementation of the same model.

\begin{figure}[t!]
\includegraphics[scale=0.625]{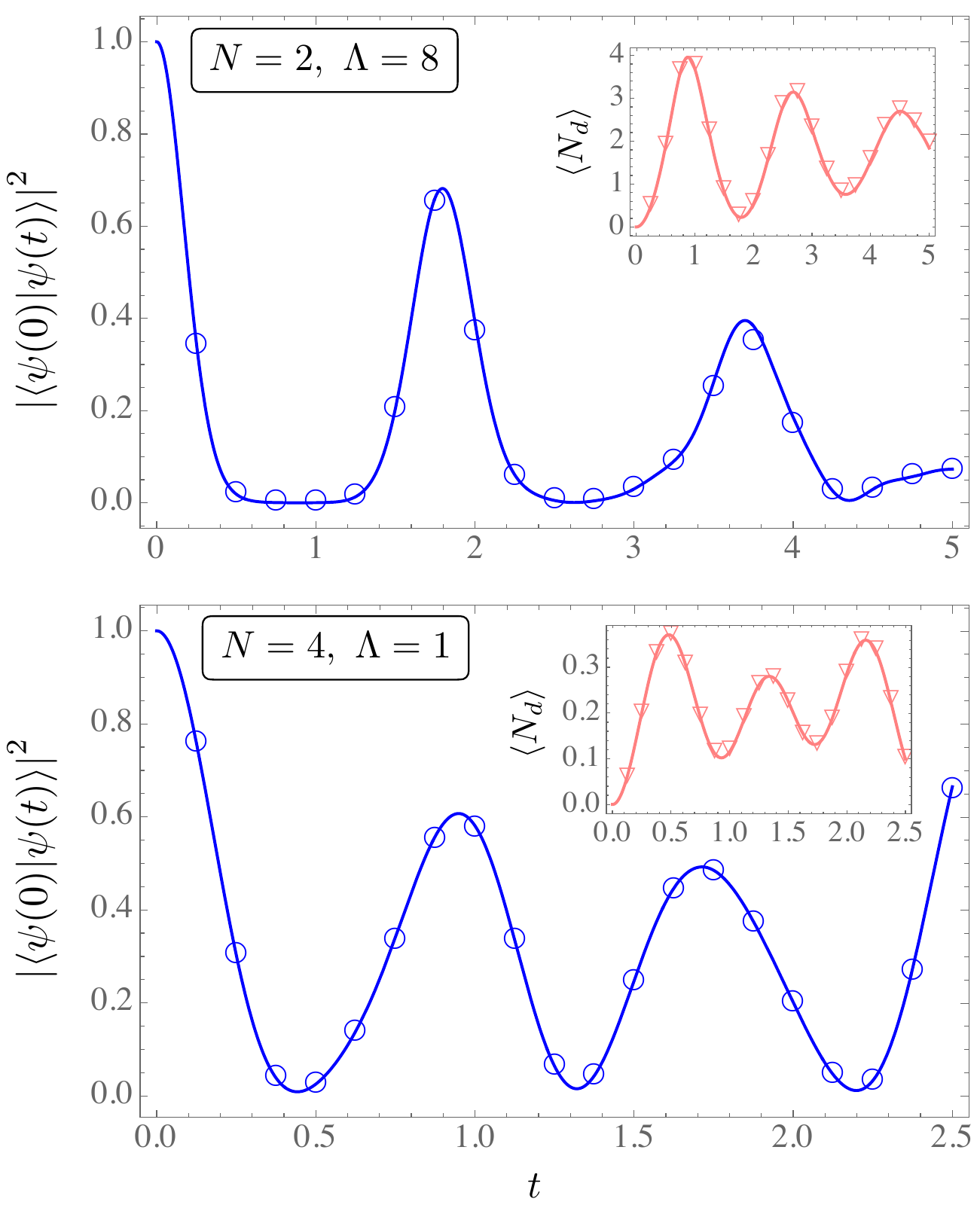}
\caption[.]{The overlap between an initial state with no fermion, no antifermion, and no boson, and the corresponding time-evolved state, $\left|\braket{\psi(0)|\psi(t)}\right|^2$, (in blue) along with the average number of bosons generated $\braket{N_d} \equiv\frac{1}{N} \braket{\psi(t)|\sum_{k=-N/2}^{N/2-1} d_k^\dagger d_k|\psi(t)}$, (in red) as a function of time, $t$, using exact (solid curve) and Trotterized evolution (unfilled points) for the Yukawa theory. The corresponding model parameters are given in Table~\ref{tab:yukawa-param}. The time $t$ is in units of lattice spacing $b$, which is set to one.
}
\label{fig:dynamic-scalar}
\end{figure}

An interesting phenomenon in the Yukawa theory considered is the dynamical generation of mass even if the bare fermion mass is set to zero originally. Such an effective mass controls the rate of fermion generation throughout the evolution and depends on the boson accumulation on the lattice. As Fig.~\ref{fig:dynamic-scalar} demonstrates for small lattice sizes $N$ and small boson occupation cutoffs $\Lambda$, such a non-trivial fermion, anti-fermion, and boson occupation evolution can be revealed in the Loschmidt echo, i.e., $|\braket{\psi(0)|\psi(t)}|^2$, where $\psi(0)$ is the state with no fermion, no antifermion, and no boson. After a quench, this state evolves to a superposition of states with any number of fermions and antifermions allowed by symmetries, and involving varying boson occupation whose average is shown in the inset of the plots. The unfilled points represent the same quantities evaluated using a Trotter expansion of the evolution operator according to the split in the Hamiltonian terms outlined in the previous section. The coarsest Trotter digitization is chosen such that quantities can still be described accurately in the time window specified.

The corresponding model parameters, shown in Table~\ref{tab:yukawa-param}, give rise to gate rotation angles denoted in Table~\ref{tab:angles-yukawa} of Appendix~\ref{app:yukawa} along with associated experimental gate parameters for the $R^{\sigma a}$ gate, including gate's operation time, in Table~\ref{tab:gate-yukawa}. The gate time required is shorter (a few tens of microseconds for reasonable experimental parameters) than that of the spin-spin entangling gate (a few hundreds of microseconds), which is a great improvement over fully-digital implementation that would have required multiple entangling spin-spin gates, being an order of magnitude slower in general. We will come back to gate-number scaling of the analog-digital implementation compared with the fully-digital implementation in Sec.~\ref{sec:cost}. The spin-spin entangling gates required in simulating the solely fermionic dynamics can be driven with the rotation angles specified in Table~\ref{tab:angles-yukawa} using well-known laser pulse-shaping techniques. The trap and laser characteristics, including transverse normal modes of motion and the Lamb-Dicke parameter, are provided in Table~\ref{tab:trap-yukawa}.

A final technical detail regarding the experimental implementation is that all normal-mode vectors must have non-zero components so that all Rabi frequencies $\Omega_j$ are finite when applying the $R^{\sigma a}$ gate. To ensure that each ion in the simulation couples to all of the phonon modes used, the total number of ions must be even. Therefore, studying an $N$-site theory of the 1+1 D Yukawa model, where the number of staggered sites $N$  is even, generally requires $N+2$ ions: one ancilla, which is operated on by the specified gates in Fig.~\ref{fig:circuit-scalar}, and one idle ion that is only present to ensure that for any given mode, no ion is stationary along the chain.

\section{U(1) lattice gauge theory coupled to (staggered) fermions in 1+1 D
\label{sec:U(1)}}
\begin{figure*}[t!]
\includegraphics[scale=0.6]{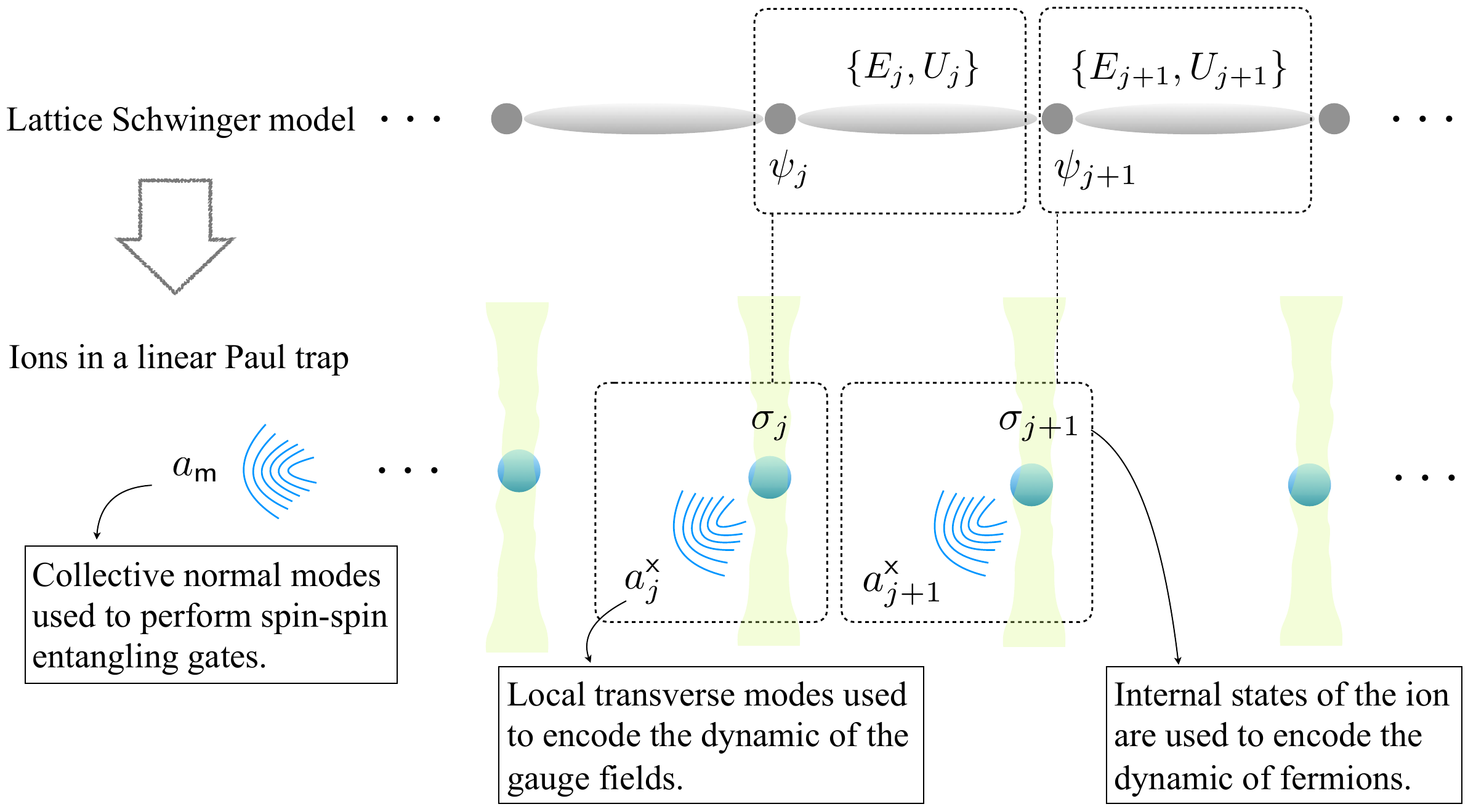}
\caption[.]{The degrees of freedom of the lattice-regularized Schwinger model in the HOBM (top row) are mapped to those in a linear trapped-ion quantum simulator involving local and normal modes of motion (bottom row).
}
\label{fig:schwinger-mapping}
\end{figure*}
\noindent
The Schwinger model, the theory of quantum electrodynamics in 1+1 D, has long served as a testbed for simulation methods (both classical and quantum) for strongly-interacting gauge theories. It is qualitatively similar to quantum chromodynamics, the theory of the strong force in nature, as it exhibits confinement, chiral-symmetry breaking, and a nontrivial $\theta$ vacuum~\cite{kogut1975hamiltonian, banks1976strong}. Being a low-dimensional and Abelian theory, it is simpler than its higher-dimensional and non-Abelian counterparts and its lattice-regularized form has been the subject of valuable quantum-hardware implementation benchmarks in recent years~\cite{Martinez:2016yna, Kokail:2018eiw, Klco:2018kyo, Lu:2018pjk, Mil:2019pbt, Yang:2020yer}. Fully-analog proposals for Hamiltonian simulation of the lattice Schwinger model have been put forward using spin degrees of freedom only~\cite{Davoudi:2019bhy}, and both the spin and phonon degrees of freedom~\cite{yang2016analog}, but they remain relatively challenging for hardware implementation. Here, we propose a hybrid analog-digital simulation of the lattice Schwinger model that brings the simulation proposals involving the phonon degrees of freedom a step closer to experimental realization.

\subsection{The Schwinger model}
The well--known Hamiltonian of the lattice Schwinger model, introduced by Kogut and Susskind~\cite{kogut1975hamiltonian, banks1976strong}, can be written in terms of a one-component staggered fermion $\psi_j$ defined on site $j$ (with odd and even sites corresponding to matter and anti-matter fields, respectively), the gauge link $U_j$ and its conjugate field, the electric field $E_j$, both defined on the link emanating from site $j$, and satisfying the commutation relations $[E_j,E_{j'}]=[U_j,U_{j'}]=0$ and $[E_j,U_{j'}]=U_j\delta_{j,j'}$. The Hamiltonian can be written as
\begin{align}
&H_{U(1)}=H_{U(1)}^{(I)}+H_{U(1)}^{(II)}+H_{U(1)}^{(III)},
\end{align}
where the gauge-matter interacting Hamiltonian, the staggered fermion-mass Hamiltonian, and the electric-field Hamiltonian are
\begin{align}
&H_{U(1)}^{(I)}=\frac{i}{2b}\sum_{j=1}^N(\psi_j^\dagger U_j^\dagger \psi_{j+1}-\psi_j U_j \psi_{j+1}^\dagger),
\\
%
%
&H_{U(1)}^{(II)}=m\sum_{j=1}^N(-1)^j\psi_j^\dagger \psi_j,
\\
%
%
&H_{U(1)}^{(III)}=\frac{g^2b}{2}\sum_{j=1}^N E_j^2,
\end{align}
respectively. The Hilbert space of the theory is characterized by on-site quantum numbers $n^{(g)}_j$ and $n^{(f)}_j$ associated with the discrete spectrum of a quantum rotor satisfying $E_j\ket{n^{(g)}_j}=n^{(g)}_j\ket{n^{(g)}_j}$ and $U_j\ket{n^{(g)}_j}=\ket{n^{(g)}_j+1}$ with $n^{(g)}_j\in \mathbb{Z}$, along with fermionic occupations eigenstates satisfying $\psi_j\ket{n^{(f)}_j}=\delta_{n^{(f)}_j,1}\ket{n^{(f)}_j-1}$ and $\psi_j^\dagger\ket{n^{(f)}_j}=\delta_{n^{(f)}_j,0}\ket{n^{(f)}_j+1}$ with $n^{(f)}_j \in \{0,1\}$. The physical states are those annihilated by the Gauss's law operator $G_j=E_j-E_{j-1}-\psi_j^\dagger \psi_j +\frac{1}{2}\big[1-(-1)^j \big]$ at each site. In the following, it is assumed that the simulation starts in a physical state and hence will remain in the same sector, as long as the gauge-symmetry violations arising from the digitization of the time-evolution operator, or from hardware imperfections, remain small.

Unfortunately, a quantum rotor representing the gauge link in the Schwinger model does not have the same algebra as the phonon modes of the trapped-ion simulator for a direct encoding. A highly-occupied boson model (HOBM) was proposed in Ref.~\cite{yang2016analog} to resolve this mismatch by considering the following mappings: $U_j \to d_j/\sqrt{M}$ and $E_j \to d_j^\dagger d_j -M$ for an integer $M$. This transformation keeps the commutation relation between $E_j$ and $U_j$ intact, but modifies that between the gauge links to $[U_j,U_{j'}]=\delta_{j,j'}/M$. Only in the limit $M \gg N$, with $N$ being the number of sites on the lattice, the HOBM will recover the lattice Schwinger model, necessitating the simulation involving the bosonic degrees of freedom to initiate in a state with a large number of bosons. Nonetheless, the numerical simulations of Ref.~\cite{yang2016analog} demonstrate that for $M \sim N$, a high level of accuracy is still achieved, particularly in space- and time-averaged dynamical observables. Applying the HOBM mapping as well as the Jordan-Wigner transformation, the spin-boson Hamiltonian of the lattice Schwinger model can be written as
\begin{align}
&H_{U(1)}'=H_{U(1)}^{(I)'}+H_{U(1)}^{(II)'}+H_{U(1)}^{(III)'},
\end{align}
where
\begin{align}
&H_{U(1)}^{(I)'}=\frac{1}{2b\sqrt{M}}\sum_{j=1}^N(\sigma_j^+d_j^{\dagger}\sigma_{i+1}^-+\sigma_{i+1}^+d_j\sigma_j^-)
\\
%
%
&\hspace{0.35 cm}=\frac{1}{8b\sqrt{M}}\sum_{j=1}^N\bigg[\sigma_j^x(d_j+d_j^\dagger)\sigma_{j+1}^x+
\sigma_j^y(d_j+d_j^\dagger)\sigma_{j+1}^y+
\nonumber
\\
&\hspace{1.65 cm}i\sigma_j^x(d_j-d_j^\dagger)\sigma_{j+1}^y-
i\sigma_j^y(d_j-d_j^\dagger)\sigma_{j+1}^x\bigg],
\label{eq:HU1Iprime}
\\
%
%
&H_{U(1)}^{(II)'}=\frac{m}{2}\sum_{j=1}^N(-1)^j\sigma^z_j+{\rm const.},
\label{eq:HU1IIprime}
\\
%
%
&H_{U(1)}^{(III)'}=\frac{g^2b}{2}\sum_{j=1}^N(d_j^\dagger d_j-M)^2
\\
%
%
&\hspace{0.65 cm}=\frac{g^2b}{2}\sum_{j=1}^N\bigg[-2M(d_j^\dagger d_j)+(d_j^\dagger d_j)^2\bigg]+{\rm const.}
\label{eq:HU1IIIprime}
\end{align}
\begin{figure*}[t!]
\includegraphics[scale=0.65]{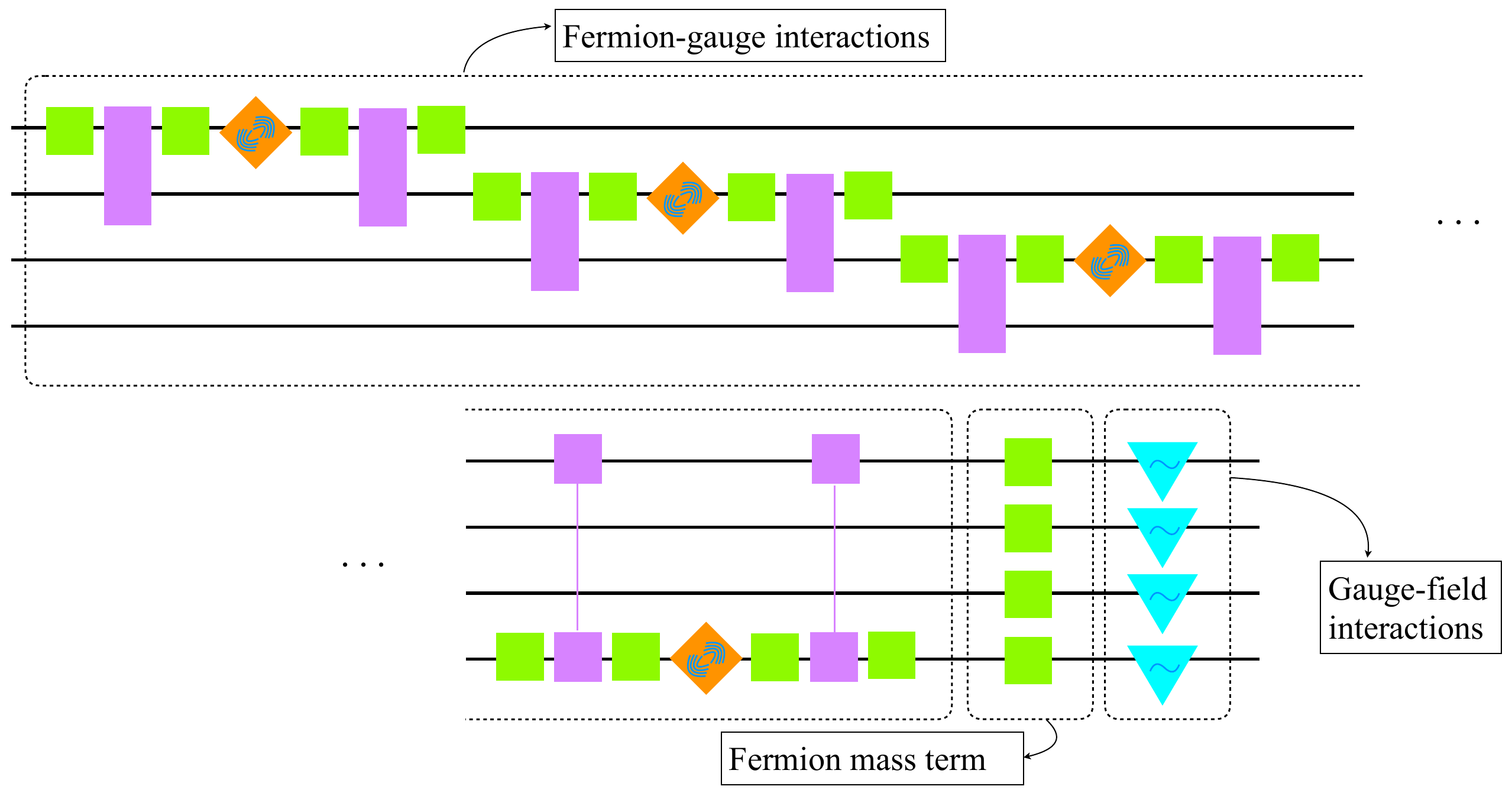}
\caption[.]{The schematic of the analog-digital quantum circuit associated with the time evolution of a four-site Schwinger model for a single Trotter step, as expressed in Eqs.~(\ref{eq:expHU1Iprime})-(\ref{eq:expHU1IIIprime}). The fermion-gauge interaction block must be repeated four times with various phases for the spin-phonon gates and single-spin rotations, see Eq.~(\ref{eq:expHU1Iprime}). The gate symbols are defined in Fig.~\ref{fig:gates}.
}
\label{fig:circuit-schwinger}
\end{figure*}
\subsection{The mapping to an analog-digital circuit}
This Hamiltonian can be directly mapped to spin and phonon degrees of freedom in the trapped-ion simulator. Noting that the electric-field Hamiltonian in the HOBM, $H_{U(1)}^{(III)'}$, consists of local self-interactions of the bosonic modes, it is necessary to consider a tight trap in the transverse directions, as introduced in Sec.~\ref{sec:local}, to take advantage of the local modes of the motion. In particular, the boson encoding should be realized by the mapping $d_j \to a_j^{\mathsf{x}}$, and the local self-interaction of the phonons can be implemented by the standing-wave gate introduced in Eq.~(\ref{eq:Raa})

The Trotterized evolution operator is $e^{-iH_{U(1)}' \delta t}=e^{-iH_{U(1)}^{(I)'}\delta t}e^{-iH_{U(1)}^{(II)'}\delta t}e^{-iH_{U(1)}^{(III)'}\delta t}+\mathcal{O}\big((\delta t)^2\big)$, where
\begin{align}
&e^{-iH_{U(1)}^{(I)'}\delta t}=\bigg[\prod_{j=1}^N S_j \mathcal{R}_{j,j+1}^{\sigma\sigma}(\frac{\pi}{4})R_j^\sigma(\frac{\pi}{4},0)\times
\nonumber
\\
%
%
&\hspace{1.4 cm}
R^{\sigma a}_j(\theta_j,0) R_j^\sigma(-\frac{\pi}{4},0) \mathcal{R}_{j,j+1}^{\sigma\sigma}(-\frac{\pi}{4})S_j^\dagger\bigg]
\nonumber\\
&\hspace{1.4 cm}
\bigg[\prod_{j=1}^N S_j^2S_{j+1}\mathcal{R}_{j,j+1}^{\sigma\sigma}(\frac{\pi}{4})R_j^\sigma(\frac{\pi}{4},0)\times
\nonumber\\
&\hspace{1.4 cm}
R^{\sigma a}_j(\theta_j,0) R_j^\sigma(-\frac{\pi}{4},0) \mathcal{R}_{j,j+1}^{\sigma\sigma}(-\frac{\pi}{4}){S^\dagger_j}^2S^\dagger_{j+1}\bigg]
\nonumber
\\
%
%
&\hspace{1.4 cm}
\bigg[\prod_{j=1}^N S_j S_{j+1}\mathcal{R}_{j,j+1}^{\sigma\sigma}(\frac{\pi}{4})R_j^\sigma(\frac{\pi}{4},0)\times
\nonumber
\\
%
%
&\hspace{1.4 cm}
R^{\sigma a}_j(\theta_j,\frac{\pi}{2}) R_j^\sigma(-\frac{\pi}{4},0) \mathcal{R}_{j,j+1}^{\sigma\sigma}(-\frac{\pi}{4})S^\dagger_{j+1}S_j^\dagger\bigg]
\nonumber
\\
%
%
&\hspace{1.4 cm}
\bigg[\prod_{j=1}^N S_j^2\mathcal{R}_{j,j+1}^{\sigma\sigma}(\frac{\pi}{4})R_j^\sigma(\frac{\pi}{4},0)\times
\nonumber
\\
%
%
&\hspace{1.4 cm}
R^{\sigma a}_j(-\theta_j,\frac{\pi}{2}) R_j^\sigma(-\frac{\pi}{4},0) \mathcal{R}_{j,j+1}^{\sigma\sigma}(-\frac{\pi}{4}){S^\dagger_{j}}^2\bigg],
\label{eq:expHU1Iprime}
\end{align}
where each square bracket corresponds to the time evolution of each of the four terms in Eq.~(\ref{eq:HU1Iprime}) and $\theta_j=\frac{1}{8b\sqrt{M}}\delta t$, 
\begin{align}
&e^{-iH_{U(1)}^{(II)'}\delta t}=\prod_{j=1}^N R_j^z(\theta'_j),
\label{eq:expHU1IIprime}
\end{align}
with $\theta'_j = \frac{1}{2} (-1)^j m \delta t$, and with the gates defined in Secs.~\ref{sec:normal}, and
\begin{align}
&e^{-iH_{U(1)}^{(III)'}\delta t}=\prod_{j=1}^NR^{aa}_{j}(\chi^{(1)},\chi^{(2)}),
\label{eq:expHU1IIIprime}
\end{align}
with $\chi^{(1)}=-g^2bM\delta t$ and $\chi^{(2)}=\frac{1}{2}g^2b\delta t$, and with the gates defined in Secs.~\ref{sec:local}. The schematic of the corresponding circuit is shown in Fig.~\ref{fig:circuit-schwinger}. Note that the MS gate $\mathcal{R}_{j,j+1}^{\sigma\sigma}$ in  Eq.~(\ref{eq:expHU1Iprime}) can be implemented using the normal modes of the motion along the axial direction, as the phonons involved in its implementation are virtual and not contributing to the dynamics of gauge fields in the simulated theory. Furthermore, the ratio $\chi^{(1)}/\chi^{(2)}$ in Eq.~(\ref{eq:expHU1IIIprime}) is fixed in the gate design as mentioned before, and one must ensure that $\chi^{(1)}/\chi^{(2)}=(-1+\widetilde{\eta}^2)/\widetilde{\eta}^2=-2M$, with the standing-wave Lamb-Dicke parameter $\widetilde{\eta}$ defined in Sec.~\ref{sec:local}. As typically $\widetilde{\eta}^2 \ll 1$, the requirement $M \gg N$ is guaranteed, but the careful tuning of $\widetilde{\eta}$ in experiment is required as the thermodynamic limit of the HOBM ($N \to \infty$) is taken. If the necessary variation in $\widetilde{\eta}$ is not feasible in a given experimental setup, one can adjust the detuning of the laser beatnotes, as explained in Sec.~\ref{sec:Yukawa}, to effectively shift the free Hamiltonian of the trapped-ion system by a term proportional to $\sum_j {a^\mathsf{x}_j}^\dagger a^\mathsf{x}_j$ in a suitable rotating frame. In this way, the corresponding term in the interaction-picture Hamiltonian, i.e., the first term in brackets in Eq.~(\ref{eq:HU1IIIprime}), can take an arbitrary coefficient, hence relaxing the condition on the ratio of ${a^\mathsf{x}_j}^\dagger a^\mathsf{x}_j$ and $({a^\mathsf{x}_j}^\dagger a^\mathsf{x}_j)^2$ terms. This will amount to changing the frequency associated with red and blue sideband transitions involved in the operation of $R^{\sigma a}_j(\theta_j,\phi_j)$ accordingly, as detailed before for the case of the Yukawa theory.

Simulating QED in higher dimensions requires engineering the magnetic Hamiltonian, that is the sum of the product of gauge links along the elementary plaquette in each two-dimensional plane in coordinate system. In the HOBM of QED, the gauge links on these higher-dimensional lattices can still be mapped to the local modes of motion in a linear chain of ions. However, engineering the magnetic Hamiltonian requires achieving four-body interactions among the phonons pinned to four adjacent ions, which will be challenging given the experimental setup explained in this work. This is because the nearest-neighbor phonon hopping needs to be simultaneously suppressed to simulate the electric-field Hamiltonian accurately. Analog and digital simulations with quantum simulators are proposed for QED in 2+1 D with pure gauge interactions, using both the HOBM or other models~\cite{Zohar:2012ts, Ott:2020ycj, Celi:2019lqy, Bender:2020ztu}. Future work will investigate the design of an analog-digital simulation approach for this problem.

\subsection{An example with realistic parameters
\label{sec:schwinger-example}}
\begin{figure}[t!]
\includegraphics[scale=0.625]{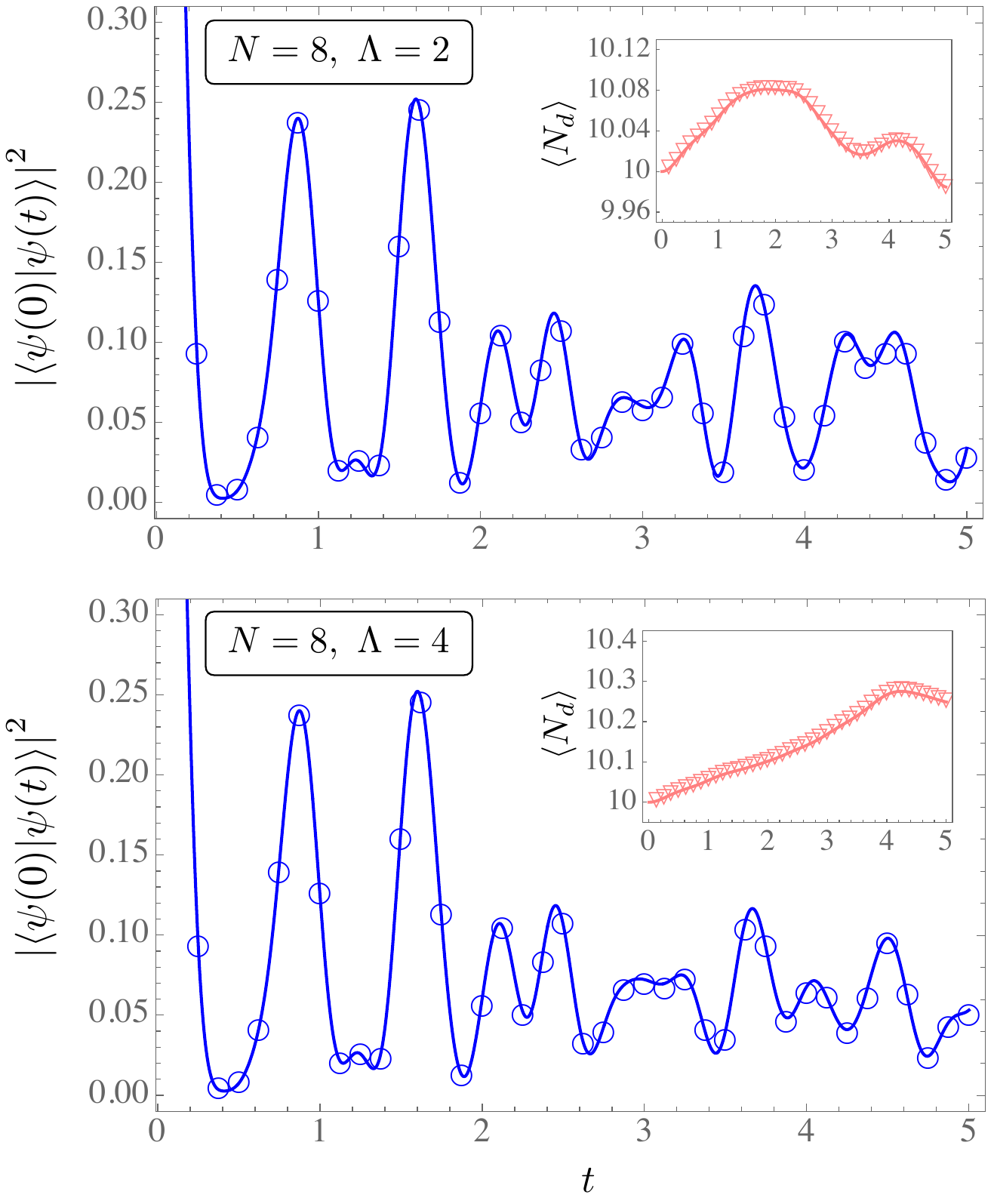}
\caption[.]{The overlap between an initial state with no fermion, no antifermion, and $M$ bosons, and the corresponding time-evolved state, $\left|\braket{\psi(0)|\psi(t)}\right|^2$, (in blue) along with the average number of phonons generated, $\braket{N_d} \equiv\frac{1}{N} \braket{\psi(t)|\sum_{j=1}^{N} d_j^\dagger d_j|\psi(t)}$, (in red) as a function of time using exact (solid curve) and Trotterized evolution (unfilled points) for the Schwinger model within HOBM. The corresponding model parameters are given in Table~\ref{tab:schwinger-param}.  The time $t$ is in units of lattice spacing $b$, which is set to one.
}
\label{fig:dynamic-schwinger}
\end{figure}

Similar to the case of the Yukawa theory, near-term experimental implementations of the Schwinger model with realistic analog-digital gate parameters will be viable for small to intermediate systems and can be scaled up straightforwardly similar to what is expected for digital implementations. It is therefore useful to consider small instances of the problem that can be simulated classically to obtain the range of expected parameters in experiment for which interesting phenomena in the model can be observed. Among many interesting features of the Schwinger model are the fermion-antifermion pair creation and the string-breaking dynamics. These quantities can be evaluated from expectation values of fermionic or bosonic observables in a state evolved from a trivial initial state, $\ket{\psi(0)}$, such as the vacuum of the model in the limit $g \to \infty$, i.e., a state with no fermion, no antifermion, and $M$ bosons. Denoting the time-evolved state at time $t$ as $\ket{\psi(t)}$, we simply consider  the Loschmidt echo, i.e., $|\braket{\psi(0)|\psi(t)}|^2$. $\ket{\psi(t)}$ involves a superposition of states with any number of fermions and antifermions allowed by symmetries, and involving varying boson occupation whose average is shown in the inset of the plots in Fig.~\ref{fig:dynamic-schwinger}. The coarsest Trotter digitization is chosen such that quantities can still be described accurately in the time window specified, and the corresponding values are marked with unfilled points. $\Lambda$ denotes the cutoff on the excitations of the bosons above the $M$ bosons in the initial state. It is notable that the boson cutoff effects are significant even for such a small lattice, motivating further the need for mapping a larger bosonic Hilbert space to the quantum simulator, which is done naturally in the analog-digital proposal of this work using the local phonon modes in the trap.
\begin{table}[b!]
\scalebox{1}
{
\begin{tabular}{ccccccccccccccc}
\hline 
\multicolumn{15}{c}{Model parameters}\tabularnewline
\hline 
\hline 
 &  &  &  &  &  &  &  &  &  &  &  &  &  & \tabularnewline
$b$ &  & $N$ &  & $\Lambda$ &  & $g$ &  & $m$ &  & $M$ &  & $\delta t$ &  & $t$\tabularnewline
 &  &  &  &  &  &  &  &  &  &  &  &  &  & \tabularnewline
\hline 
 &  &  &  &  &  &  &  &  &  &  &  &  &  & \tabularnewline
\multirow{2}{*}{1} &  & \multirow{2}{*}{8} &  & 2 &  & \multirow{2}{*}{0.35} &  & \multirow{2}{*}{1} &  & \multirow{2}{*}{10} &  & \multirow{2}{*}{0.125} &  & \multirow{2}{*}{5}\tabularnewline
 &  &  &  & 4 &  &  &  &  &  &  &  &  &  & \tabularnewline
 &  &  &  &  &  &  &  &  &  &  &  &  &  & \tabularnewline
\hline 
\end{tabular}
}
\caption{The parameters of the Schwinger model within the HOBM considered in the examples of this section. The corresponding trap and gate parameters are tabulated in Appendix~\ref{app:schwinger}. $t$ is the total evolution time and $\delta t$ is the duration of the Trotter-evolution segment.}
\label{tab:schwinger-param}
\end{table}

Since the trap is in the tight-binding regime along the transverse directions, the local-mode frequencies at the location of each ion are equal to a good approximation, simplifying the parameters of the gates applied to each ion. For the phonon-phonon gate, it is ensured that the parameter $F\widetilde{\eta}^2/\omega^\mathsf{x}$ is small by choosing $\widetilde{\eta}$ and $F$ appropriately. Furthermore, to generate a specific value of the ratio of the phonon terms ${a{^\mathsf{x}}}^\dagger a^\mathsf{x}$ and $({a^{\mathsf{x}}}^\dagger a^\mathsf{x})^2$ as required by the HOBM, the red and blue sideband transitions in the spin-spin and spin-phonon gates must be implemented with a frequency shift $-\delta \omega^\mathsf{x}$, corresponding to the necessary modification to the interaction-picture Hamiltonian. The shift value, along with other relevant trap parameters, are tabulated in Table~\ref{tab:trap-schwinger} of Appendix~\ref{app:schwinger}. As noted in Tables~\ref{tab:gate-I-schwinger} and \ref{tab:gate-II-schwinger}, the spin-phonon gate time is similar to what was obtained for the Yukawa theory. The phonon-phonon gate time needs to be $\sim 100$ microseconds to prevent the need for a large value of the standing wave amplitude. Last but not least, the Lamb-Dicke parameters associated with the Raman beams as well as the standing-wave beam must satisfy $\eta \sqrt{M} \ll 1$ and $\widetilde{\eta}\sqrt{M} \ll 1$, respectively. This is because the system must be prepared in a state with phonon occupation $M$ according to the HOBM. These conditions are satisfied with the realistic experimental parameters chosen.
%
\section{A qualitative cost analysis
\label{sec:cost}}
\noindent
In this section, our hybrid analog-digital approach will be compared with the fully-digital realization of the Yukawa theory and the Schwinger model.
In the long term, when fault-tolerant digital quantum computation will be a reality, the non-Clifford gate (T-gate) count determines the cost of the simulation. In the near-term and particularly over the next decade, however, quantum computations will be limited by the slow and error-prone application of entangling (CNOT) gates. In the analog-digital protocol of this work, while the hardware-specific spin-phonon gate is an entangling gate in the combined Hilbert space of the phonons and qubits, its operation is governed by dynamics that occur at $\mathcal{O}(\eta)$, where $\eta$ is the Lamb-Dicke parameter introduced in Sec.~\ref{sec:simulator}. This is in contrast to the entangling spin-spin operation that is the core of the CNOT implementation in a trapped-ion digital computer, which is governed by dynamics that occur at $\mathcal{O}(\eta^2)$. Since $\eta \ll 1$ in the Lamb-Dicke regime in which experiments operate, the spin-spin gates are one or two orders of magnitude slower than single-spin and spin-phonon gates. Phonon-phonon self-interaction gates based on standing-wave beams are governed by $\mathcal{O}(\widetilde{\eta}^4)$ dynamics but their strength can be compensated by laser power, as demonstrated by the examples studied before. Without experimental realizations, the fidelity of this new set of gates cannot be accurately predicted. Nonetheless, the qualitative expectations for the gate times as described here can provide a rough guidance to the relative performance of the simulator in an analog-digital mode compared with a fully-digital mode.

\subsection{The Yukawa theory
\label{sec:cost-yukawa}}
Since the implementation of the dynamics associated with the fermionic hopping and mass is the same in both digital and analog-digital circuits, here we focus on implementing dynamics involving the scalar field, namely evolution with the free scalar-field Hamiltonian as well as the fermion scalar-field interacting Hamiltonian.

In the analog-digital protocol, the time evolution of the free scalar-field Hamiltonian $H_{\text{Yukawa}}^{(II)}$ in Eq.~(\ref{eq:HYII2}) is implemented at no cost since it amounts to adjusting laser beatnote frequencies in the sideband operations, as explained in Sec.~\ref{sec:Yukawa-mapping}. In the fully-digital implementation, the exact circuit design depends on the mapping of the scalar fields to qubit registers. The field basis~\cite{Jordan:2011ci}, the harmonic-oscillator basis~\cite{Klco:2018zqz}, and the single-particle basis~\cite{Barata:2020jtq} are among the representations considered in literature to perform such a mapping, each with their own benefits and disadvantages. Here, we give a qualitative CNOT-gate count (or in turn MS entangling-gate count) in the harmonic-oscillator basis, as it is the representation closely related to what is considered in Sec.~\ref{sec:Yukawa}. In this basis, the occupation number of each harmonic oscillator corresponding to each momentum mode is truncated at some cutoff value $\Lambda$ and is expressed in a binary representation, which is then mapped to $\log \Lambda$ qubit registers. The evolution operator $e^{-iH_{\text{Yukawa}}^{(II)}\delta t}=e^{-i \sum_k \varepsilon_k \, (d^\dagger_k d_k+\frac{1}{2}) \delta t}$ can then be realized by a number of single-qubit rotations around the $z$ axis of the qubits' Bloch sphere (see e.g., Ref.~\cite{Shaw:2020udc} for a similar example of implementing the time evolution of the electric-field Hamiltonian in the lattice Schwinger model). This step, therefore, can be considered cost-free in both digital and analog-digital implementations.
\begin{table*}[t!]
\begin{tabular}{ccccc}
\hline 
\multicolumn{5}{c}{Yukawa theory }\tabularnewline
\hline 
\hline 
 & ~~Fermion hopping~~ & ~~Fermion mass~~ & ~~Free scalar fields~~ & ~~Fermion scalar-field interaction~~\tabularnewline
\hline 
Analog-digital & $\mathcal{O}\left(N\right)$ & $\mathcal{O}\left(1\right)$ & $\mathcal{O}\left(1\right)$ & $\mathcal{O}\left(1\right)$\tabularnewline
\hline 
Digital & $\mathcal{O}\left(N\right)$ & $\mathcal{O}\left(1\right)$ & $\mathcal{O}\left(1\right)$ & $\mathcal{O}\left(N^{2}\left(\log\Lambda\right)^{2}\right)$\tabularnewline
\hline 
\end{tabular}

$\vphantom{}$

$\vphantom{}$

$\vphantom{}$

\begin{tabular}{cccc}
\hline 
\multicolumn{4}{c}{Schwinger model }\tabularnewline
\hline 
\hline 
 & ~~Fermion-gauge interaction~~ & ~~Fermion mass~~ & ~~Electric-field term~~\tabularnewline
\hline 
Analog-digital & $\mathcal{O}\left(N\right)$ & $\mathcal{O}\left(1\right)$ & $\mathcal{O}(1)$\footnote{$\mathcal{O}(N)$ standing-wave phonon-phonon gates which are expected to have comparable fidelity to the spin-spin gates.}\tabularnewline
\hline 
Digital & $\mathcal{O}\left(N\left(\log\Lambda\right)^{2}\right)$ & $\mathcal{O}\left(1\right)$ & $\mathcal{O}\left(N\left(\log\Lambda\right)^{2}\right)$\tabularnewline
\hline 
\end{tabular}
\caption{The scaling of the spin-spin gate count per Trotter step as a function of the lattice size $N$ and the cutoff on the boson (phonon) excitations $\Lambda$ for the Yukawa theory and the Schwinger model (within HOBM) assuming analog-digital and fully-digital implementations. The spin-phonon and phonon-phonon gate counts are not denoted in the table for the analog-digital approach but have been discussed in the text.}
\label{tab:cost}
\end{table*}

Next, the time evolution with the fermion scalar-field Hamiltonian $H_{\text{Yukawa}}^{(III)}$ in Eq.~(\ref{eq:HYIII}) is implemented with a single ancilla qubit and with $\mathcal{O}(N(N+1))$ spin-phonon gates, which are considered to be cost-free compared with the entangling spin-spin gates. In a fully-digital implementation, $e^{-iH_{\text{Yukawa}}^{(III)}\delta t}$ can be implemented by writing $H_{\text{Yukawa}}^{(III)}$ as
\begin{align}
&H_{\text{Yukawa}}^{(III)}=
\sqrt{\frac{g^2b}{2N}} \sum_j\sum_k \sqrt{\frac{1}{\varepsilon_k}}\cos\big(\frac{2\pi kj}{N}\big) \psi^\dagger_j\psi_j (d_k+d_k^\dagger)+
\nonumber\\
&\hspace{1.6 cm} \sqrt{\frac{g^2b}{2N}} \sum_j\sum_k \sqrt{\frac{1}{\varepsilon_k}}\sin(\frac{2\pi kj}{N}) \psi^\dagger_j\psi_j i(d_k-d_k^\dagger).
\label{eq:HYIII-split}
\end{align}
The implementation of these terms follows the procedure explained in Ref.~\cite{Shaw:2020udc} in the case of realizing the dynamics of fermion-gauge interaction term in the Schwinger model. First, Eq.~(\ref{eq:HYIII-split}) indicates that the operations proportional to $A \equiv (d_k+d_k^\dagger)$ and $B \equiv i(d_k-d_k^\dagger)$ operators are only performed if the fermion occupies a given site, necessitating a controlled operation on the qubit register of the fermion. The $A$ and $B$ operators are two near-diagonal matrices whose exponential can be implemented using the shift operators, that are realized using quantum Fourier transform circuits and single-qubit rotations in the Fourier space. As a reminder, for a binary number with $\log \Lambda$ digits, each quantum Fourier transform requires $\mathcal{O}((\log \Lambda)^2+\log \Lambda)$ CNOT operations. Relating $A$ and $B$ operators to the shift operators requires a periodic wrapping of the matrices, i.e., identifying the least and most values of harmonic-oscillator occupation. This unphysical modification can subsequently be removed by application of appropriate $\log \Lambda$-controlled operations, amounting to $\mathcal{O}(\log \Lambda)$ additional CNOT gates~\cite{Shaw:2020udc}. Putting everything together, including the controlled operations required on the fermionic register, and taking into account all the terms in the lattice sums in Eq.~(\ref{eq:HYIII-split}), the time evolution operator $e^{-iH_{\text{Yukawa}}^{(III)}\delta t}$ can be implemented using $\mathcal{O}(N^2(\log \Lambda)^2)$ CNOT operations. Therefore, assuming that spin-phonon gates of the hybrid scheme are free compared with spin-spin entangling gates, the digital approach is inferior to the hybrid approach. Even if the spin-phonon gate performs comparably to the spin-spin (CNOT) gate, the digital scheme require $\mathcal{O}((\log \Lambda)^2)$ more entangling operations which can be significant when $\Lambda \gg 1$.

Such an advantage is at the core of the power of the hybrid approach: phonons are represented naturally and as many phonon excitations as permitted in the dynamics can be generated without the need to cut their spectrum off. Of course, an excessive number of phonons in the system can lead to Kerr cross-coupling \cite{Marquet2003} and loss of coherence in the simulator, and therefore a balance should be established between accuracy of the simulated theory given a truncated boson spectrum and the experimental error in the simulator. For this reason, experimental benchmarks are necessary in confirming these qualitative theoretical expectations.  A summary of the entangling-gate count of both schemes for evolving each term in the Hamiltonian of the Yukawa theory is provided in Table~\ref{tab:cost}.

\subsection{The Schwinger model
\label{sec:cost-schwinger}}
Except for the time evolution of the fermion mass term, both the fermion-gauge field interaction and the electric-field term (the boson self-interactions in the HOBM) are implemented differently in the hybrid and fully-digital schemes.

The circuit in Fig.~\ref{fig:circuit-schwinger} reveals that the time evolution of the interacting fermion-boson field in the HOBM requires $\mathcal{O}(N)$ spin-spin gates and $\mathcal{O}(N)$ spin-phonon gates, with the latter expected to be not too costly. In the fully-digital scheme, $e^{-iH_{U(1)}^{(I)'}\delta t}$ with the Hamiltonian in Eq.~(\ref{eq:HU1Iprime}) can be implemented following the circuit construction described earlier in the case of fermion scalar-field interactions of the Yukawa theory. The only differences are that now the bosons are defined locally, and associated with each site (link) there is one such boson (as opposed to $N$ bosons associate with all momentum modes in the Yukawa theory), and that the fermions correspond to nearest-neighbor sites (as opposed to a local fermion occupation operator in the corresponding term in the Yukawa theory). As a result, the total entangling-gate count of the digital-circuit implementation of this evolution operator is $\mathcal{O}(N(\log \Lambda)^2)$, where $\Lambda$ denotes the cutoff on the boson excitations (translating to the electric-field excitations in the original U(1) theory). Therefore, the hybrid implementation will save a factor of $\mathcal{O}((\log \Lambda)^2)$ in CNOT count compared with the digital implementation assuming again that the spin-phonon gate performs at a much higher fidelity than the spin-spin gate.

Time evolving the electric-field term, that is Eq.~(\ref{eq:HU1IIIprime}), requires $N$ standing-wave gates, which can be counted to be comparable to $\mathcal{O}(N)$ entangling gates. The same term in the fully-digital computation requires either free Z-rotations (for evolving the $d^\dagger_j d_j$ term similar to what was discussed for the free scalar Hamiltonian in the Yukawa theory) or $\mathcal{O}(N\log \Lambda (\log \Lambda-1))$ CNOT gates (for evolving the $(d^\dagger_j d_j)^2$ term). This latter count can be understood by noting that the application of $d^\dagger_j d_j$ phase depends on the occupation of the $d$ excitation itself. In summary, the analog-digital implementation requires $\mathcal{O}((\log \Lambda)^2)$ fewer entangling operations (assuming that the standing-wave gate can perform comparably to the spin-spin gate). A summary of the entangling-gate count of both schemes for evolving each term in the Hamiltonian of the HOBM is provided in Table~\ref{tab:cost}.

\section{Conclusion and outlook
\label{sec:conclusion}}
\noindent
Our work combines the benefits of analog and digital trapped-ion simulators to enable both short- and long-term simulations of quantum field theories involving bosonic fields with sizable Hilbert spaces. Among the models considered are the Yukawa theory, i.e., scalar fields coupled to fermions in 1+1 D and the lattice Schwinger model within the highly-occupied bosonic model. By  introducing a set of known and new gates, including spin, spin-spin, spin-phonon, and phonon-phonon gates, the Trotterized evolution in each of these models is expressed as a quantum circuit. Small instances of the models with realistic experimental parameters for near-term proof-of-principle demonstrations of non-trivial dynamics are identified. Experiments with 20 ions using a set of normal or local modes, each occupied with only up to two phonon excitations, already push the limits of what is possible on a classical computer. Most importantly, higher cutoffs on the bosonic excitations can be imposed in the model as the bosons are naturally represented by the phonons in the ion trap, leading to an $\mathcal{O}\left(N \log \Lambda\right)$ ($\mathcal{O}\left(\log \Lambda\right)$) reduction in the number of qubits and at least an $\mathcal{O}\left((N^2\log \Lambda)^2\right)$ ($\mathcal{O}\left((\log \Lambda)^2\right)$) reduction in the number of the entangling spin-spin gates compared with a fully-digital simulation in the Yukawa theory (Schwinger model). Here, $\Lambda$ is the cutoff on the boson-field excitation and $N$ is the number of lattice sites in the simulated theory. These scalings assume that the spin-phonon gate has a higher fidelity compared with the spin-spin gate, while the phonon-phonon gate performs with a fidelity comparable to that of a spin-spin gate.  Experimental implementations will test these theoretical expectations in the coming years and establish the analog-digital mode of the simulator as a more powerful paradigm than either of fully-analog and digital schemes in enabling simulations of interacting fermionic-bosonic field theories of phenomenological interest.

A number of promising theoretical and experimental directions may result from extensions of this work to other quantum simulators and more complex strongly-interacting field theories. These include the following:

\itemize

\item[$\diamond$]
The bosonic modes are common in a range of quantum simulators. A notable example is a circuit-QED platform in which cavity photons are coupled to superconducting qubits~\cite{blais2007quantum, schmidt2013circuit}. The physics of photon-qubit effective interactions is very similar to that of phonon-ion interactions in an ion trap. For example, cavity photons can be taken advantage of to induce non-local couplings among the qubits~\cite{zhu2018hardware}. One can envision encoding the dynamics of scalar or gauge fields in the photon modes and that of fermions in the superconducting qubits, hence devising photon-based gates as discussed in this work. Nonetheless, the degree of control and the feasibility of experimental realizations must be examined carefully before the potential of an analog-digital scheme can be evaluated in such platforms.

\item[$\diamond$]
A highly desirable capability, that was not required for the current work given the nature of the models studied, is the engineering of non-local phonon-phonon couplings. Such a capability would allow the simulations of a self-interacting scalar field theory, as well as the plaquette (magnetic) interactions in lattice gauge theories in higher dimensions, in an analog-digital fashion. It will be worth exploring the possibility of taking advantage of the virtual spins to mediate the phonon interactions in this context, inspired by the M\o lmer-S\o rensen scheme, in which spin-spin interactions are mediated by the virtual phonons. Promising experimental demonstrations in circuit-QED have emerged~\cite{gao2018programmable} and similar ideas may be applicable to the trapped-ion systems.

\item[$\diamond$]
To make progress toward the goal of simulating quantum chromodynamics, one must develop feasible simulation proposals for non-Abelian gauge theories. The SU(2) LGT coupled to fermions is a valuable non-Abelian gauge theory that has become the focus of intense studies in recent years~\cite{Zohar:2012xf, Zohar:2013zla, Zohar:2014qma, Mezzacapo:2015bra, Zohar:2019ygc, Klco:2019evd, Davoudi:2020yln, Dasgupta:2020itb, Atas:2021ext, Rahman:2021yse}. Its prepotential formulation in terms of Schwinger bosons~\cite{Raychowdhury:2019iki} appears the most suitable representation to potentially take advantage of the analog-digital setup presented here. Basically, despite the U(1) theory in 1+1 D which has only an approximate representation in terms of bosonic degrees of freedom, the SU(2) LGT in any dimension has an exact mapping to bosons. The complication is that each link on the lattice is a collection of four sets of oscillators and the physical degrees of freedom are those made up of gauge-invariant combinations of  these bosonic operators (along with the fermionic operators). The Hamiltonian matrix elements retain the knowledge of the SU(2) algebra and hence contain non-trivial factors~\cite{Raychowdhury:2019iki}. Ongoing progress in developing fully-digital algorithms for this theory, nonetheless, will be beneficial in developing the analog-digital counterparts. Here, complications associated with the non-local phonon-phonon interactions in the trapped-ion simulator must be dealt with as well.

\item[$\diamond$]
It will be interesting to consider efficient state-preparation strategies for QFTs when bosonic fields can be encoded into bosonic registers. Besides the reduction in the number of qubits and entangling gates required for the preparation routine in an analog-digital mode, one can envision more efficient state-preparation protocols as well, such that non-trivial initial states involving entangled boson and fermion modes can be achieved more straightforwardly. This direction is, however, less developed and requires additional investigation.

\item[$\diamond$]
Last but not least, it is important to investigate whether the experimental imperfections and non-ideal gate fidelities are more significant in an analog-digital compared to a fully-digital setting. This is a rather crucial question when it comes to gauge-theory simulations, as local gauge invariance imposed on an initial state needs to be efficiently retained throughout the evolution, or the simulated physics will be fundamentally different than the desired gauge theory. Recent ideas in protecting gauge invariance in the simulation~\cite{Tran:2020azk, Lamm:2020jwv, Halimeh:2020ecg, kasper2020non} can be investigated in the context of analog-digital simulations as well.

\section*{Acknowledgments}
\noindent
We acknowledge valuable discussions with Mohammad Hafezi and Alireza Seif. ZD is supported in part by the U.S. Department of Energy's Office of Science Early Career Award, under award no. DE-SC0020271, for theoretical developments for mapping QFTs to quantum simulators, and by the DOE Office of Science, Office of Advanced Scientific Computing Research (ASCR) Quantum Computing Application Teams program, under fieldwork proposal number ERKJ347, for algorithmic developments for scientific applications of near-term quantum hardware. GP and NML acknowledge support by the DOE Office of Science, Office of Nuclear Physics, under Award no. DE-SC0021143, for designing hardware-specific simulation protocols for applications in nuclear physics. 
GP and NML are further supported by the Army Research Office (W911NF21P0003) and the Office of Naval Research (N00014-20-1-2695).

\bibliography{bibi.bib}

\appendix
\section{Details of the numerical examples
\label{app:details}}
\noindent
This appendix supplements the numerical examples provided in Secs.~\ref{sec:yukawa-example} and \ref{sec:schwinger-example} of the main text.
\subsection{Yukawa theory
\label{app:yukawa}}
This section contains the numerical values for the gate rotation angles (Table~\ref{tab:angles-yukawa}) and gate parameters (Table~\ref{tab:gate-yukawa}) required in implementing the examples studied in Sec.~\ref{sec:yukawa-example}. The relevant trap and laser parameters for a realistic near-term experiment are also provided (Table~\ref{tab:trap-yukawa}). 
\begin{table*}[t!]
\scalebox{0.8}
{
\begin{tabular}{ccccc}
\hline 
\multicolumn{5}{c}{Gate angles}\tabularnewline
\hline 
\hline 
 &  &  &  & \tabularnewline
$N_{\mathrm{ion}}$ & $\theta_{j,j+1}$ & $\theta_{j}$ & $\theta_{\mathsf{m},j}=\theta_{m,N+1}$ & $\phi_{\mathsf{m},j}$\tabularnewline
 &  &  &  & \tabularnewline
\hline 
 &  &  &  & \tabularnewline
2+1 & $\left\{ 0.0625,0.0625\right\} $ & $\left\{ -0.125,0.125\right\} $ & $\left\{ \left\{ 0.344,0.344,0.344\right\} ,\left\{ 0.625,0.625,0.625\right\} \right\} $ & $\{\{-\pi,-2\pi\},\{0,0\}\}$\tabularnewline
 &  &  &  & \tabularnewline
\hline 
 &  &  &  & \tabularnewline
\multirow{4}{*}{4+1} & \multirow{4}{*}{$\left\{ 0.0312,0.0312,0.0312,0.0312\right\} $} & \multirow{4}{*}{$\left\{ -0.0625,0.0625,-0.0625,0.0625\right\} $} & $\left\{ \left\{ 0.061,0.061,0.061,0.061\}\right\} ,\right.$ & $\left\{ \{-\pi,-2\pi,-3\pi,-4\pi,-5\pi\right\} ,$\tabularnewline
 &  &  & $\left\{ 0.081,0.081,0.081,0.081\right\} ,$ & $\left\{ -\pi/2,-\pi,-3\pi/2,-2\pi,-5\pi/2\right\} ,$\tabularnewline
 &  &  & $\left\{ 0.110,0.110,0.110,0.110\right\} ,$ & $\left.\left\{ 0,0,0,0,0\right\} ,\left\{ \pi/2,\pi,3\pi/2,2\pi,5\pi/2\right\} \right\} $\tabularnewline
 &  &  & $\left.\left\{ 0.081,0.081,0.081,0.081\right\} \right\} $ & \tabularnewline
 &  &  &  & \tabularnewline
\hline 
\end{tabular}
}
\caption{The gate angles associated with the circuit implementation of the Trotterized dynamics in the Yukawa theory as specified in Eqs.~(\ref{eq:expHYIprime})-(\ref{eq:expHYIVprime}). The values are associated with the model parameters given in Table~\ref{tab:yukawa-param} of the main text.}
\label{tab:angles-yukawa}
\end{table*}
\begin{table*}[t!]
\scalebox{0.985}
{
\begin{tabular}{cccc}
\hline 
\multicolumn{4}{c}{Implementing $R_{\mathsf{\left\{ m\right\} },j}^{\sigma a}(\left\{ \theta_{\mathsf{m},j}\right\} ,\left\{ \phi_{\mathsf{m},j}\right\} )$}\tabularnewline
\hline 
\hline 
 &  &  & \tabularnewline
$N_{\mathrm{ion}}$ & $\eta_{\mathsf{m},j}$ & $\frac{\varOmega_{\mathsf{m},j}}{2\pi}$ {[}kHz{]} & $\tau_{\mathrm{gate}}^{\sigma a}$ {[}ms{]}\tabularnewline
 &  &  & \tabularnewline
\hline 
\multirow{4}{*}{$2+1\:(\mathsf{m}=1,3,j=1,2,3)$} &  &  & \multirow{4}{*}{$20\times10^{-3}$}\tabularnewline
 & $\left\{ \left\{ -0.039,-0.039,-0.039\right\} ,\right.$ & $\left\{ \left\{ 98.6,98.6,98.6\right\} ,\right.$ & \tabularnewline
 & $\left.\left\{ 0.028,-0.057,0.028\right\} \right\} $ & $\left.\left\{ -248.4,124.2,-248.4\right\} \right\} $ & \tabularnewline
 &  &  & \tabularnewline
\hline 
\multirow{6}{*}{$4+2\:(\mathsf{m}=1,\cdots,4,j=1,\cdots,5)$} &  &  & \multirow{6}{*}{$10\times10^{-3}$}\tabularnewline
 & $\left\{ \left\{ -0.028,-0.028,-0.028,-0.028,-0.028\right\} ,\right.$ & $\left\{ \left\{ 69.7,69.7,69.7,69.7,69.7\right\} ,\right.$ & \tabularnewline
 & $\left\{ -0.042,-0.023,-0.008,0.008,0.023\right\} ,$ & $\left\{ 61.8,109.5,336.3,-336.3,-109.5\right\} ,$ & \tabularnewline
 & $\left\{ 0.038,-0.009,-0.029,-0.029,-0.009\right\} ,$ & $\left\{ -91.7,380.6,120.8,120.8,380.6\right\} ,$ & \tabularnewline
 & $\left.\left\{ 0.025,-0.038,-0.0196,0.0196,0.038\right\} \right\} $ & $\left.\left\{ -102.3,67.4,131.8,-131.8,-67.4\right\} \right\} $ & \tabularnewline
 &  &  & \tabularnewline
\hline 
\end{tabular}
}
\caption{The spin-phonon gate parameters associated with the gate rotation angles specified in Table~\ref{tab:angles-yukawa} for the examples studied in Sec.~\ref{sec:yukawa-example}.}
\label{tab:gate-yukawa}
\end{table*}
\begin{table*}[t!]
\begin{tabular}{cccc}
\hline 
\multicolumn{4}{c}{Normal-mode spectrum and interaction-picture shift}\tabularnewline
\hline 
\hline 
 &  &  & \tabularnewline
$N_{\mathrm{ion}}$ & $\eta$ &  $\frac{1}{2\pi}\omega_{\mathsf{m}}\:[\mathrm{kHz}]$ & $\frac{1}{2\pi}\frac{\widetilde{\varepsilon}_{\mathsf{m}}}{N+1}\:[\mathrm{kHz}]$\tabularnewline
 &  &  & \tabularnewline
\hline 
 &  &  & \tabularnewline
2+1 & 0.068 & $\left\{ 4000,3938.3,3850.2\right\} $ & $\left\{ 2.2,0, 0.7\right\} $\tabularnewline
 &  &  & \tabularnewline
\hline 
 &  &  & \tabularnewline
4+2 & 0.068 & $\left\{ 4000,3938.3,3849.4,3735.5,3596.4,3430.7\right\} $ & $\left\{ 1.3, 0.7, 0.4, 0.7,0,0\right\} $\tabularnewline
 &  &  & \tabularnewline
\hline
\end{tabular}
\caption{The characteristics of the ion trap and the lasers relevant for the examples studied in Sec.~\ref{sec:yukawa-example} and the associated gate parameters in Table~\ref{tab:gate-yukawa}.}
\label{tab:trap-yukawa}
\end{table*}
\subsection{Schwinger model
\label{app:schwinger}}
This section contains the numerical values for the gate rotation angles (Table~\ref{tab:angles-schwinger}) and gate parameters (Tables~\ref{tab:gate-I-schwinger} and \ref{tab:gate-II-schwinger}) required in implementing the examples studied in Sec.~\ref{sec:schwinger-example}. The relevant trap and laser parameters for a realistic near-term experiment are also provided (Table~\ref{tab:trap-schwinger}). 
\begin{table*}[t!]
\begin{tabular}{ccccccc}
\hline 
\multicolumn{7}{c}{Gate angles}\tabularnewline
\hline 
\hline 
 &  &  &  &  &  & \tabularnewline
$\theta_{j}$ &  & $\theta_{j}'$ &  & $\chi^{\left(1\right)}$ &  & $\chi^{\left(2\right)}$\tabularnewline
 &  &  &  &  &  & \tabularnewline
\hline 
 &  &  &  &  &  & \tabularnewline
$0.008$ for all $j$ &  & $\left\{ -0.0625,0.0625,-0.0625,0.0625,-0.0625,0.0625,-0.0625,0.0625\right\} $ &  & $-0.153$ &  & $0.008$\tabularnewline
 &  &  &  &  &  & \tabularnewline
\hline 
\end{tabular}
\caption{The gate rotation angles associated with the circuit implementation of the Trotterized dynamics in the Yukawa theory as specified in Eqs.~(\ref{eq:expHU1Iprime})-(\ref{eq:expHU1IIIprime}). The values are associated with the model parameters given in Table~\ref{tab:schwinger-param} of the main text.}
\label{tab:angles-schwinger}
\end{table*}
\begin{table}[t!]
\begin{tabular}{ccc}
\hline 
\multicolumn{3}{c}{Implementing $R_{j}^{\sigma a}(\theta_{j},\phi)$}\tabularnewline
\hline 
\hline 
 &  & \tabularnewline
$\varOmega_{j}/2\pi$ {[}kHz{]} &  & $\tau_{\mathrm{gate}}^{\sigma a}$ {[}ms{]}\tabularnewline
 &  & \tabularnewline
\hline 
 &  & \tabularnewline
$-43.9$ for all $j$ &  & $10^{-3}$\tabularnewline
 &  & \tabularnewline
\hline 
\end{tabular}
\caption{The spin-phonon gate parameters associated with the gate rotation angles specified in Table~\ref{tab:angles-schwinger} for the examples studied in Sec.~\ref{sec:schwinger-example}. The required $\phi$ angles are already noted in the relevant circuit expression in Eq.~(\ref{eq:expHU1Iprime}).}
\label{tab:gate-I-schwinger}
\end{table}
\begin{table}[t!]
\begin{tabular}{ccccccccc}
\hline 
\multicolumn{9}{c}{Implementing $R_{j}^{aa}(\chi^{\left(1\right)},\chi^{\left(2\right)})$}\tabularnewline
\hline 
\hline 
 &  &  &  &  &  &  &  & \tabularnewline
$F$/2$\pi$ {[}kHz{]} &  & $\widetilde{\eta}$ &  & $\widetilde{\eta}\sqrt{M}$ &  & $F\widetilde{\eta}^{2}/\omega^{\mathsf{x}}$ &  & $\tau_{\mathrm{gate}}^{aa}$ {[}ms{]}\tabularnewline
 &  &  &  &  &  &  &  & \tabularnewline
\hline 
 &  &  &  &  &  &  &  & \tabularnewline
1949.6 &  & $0.05$ &  & 0.16 &  & $0.001$ &  & $50\times10^{-3}$\tabularnewline
 &  &  &  &  &  &  &  & \tabularnewline
\hline 
\end{tabular}
\caption{The phonon-phonon gate parameters associated with the gate rotation angles specified in Table~\ref{tab:angles-schwinger} for the examples studied in Sec.~\ref{sec:schwinger-example}.
}
\label{tab:gate-II-schwinger}
\end{table}
\begin{table}[t!]
\begin{tabular}{ccccccccc}
\hline 
\multicolumn{9}{c}{Local-mode properties and interaction-picture shift}\tabularnewline
\hline 
\hline 
 &  &  &  &  &  &  &  & \tabularnewline
$N_{\mathrm{ion}}$ &  & $\frac{1}{2\pi}\omega^{\mathsf{x}}$ {[}kHz{]} &  & $\eta_{j}\equiv\eta$ &  & $\eta\sqrt{M}$ &  & $\frac{1}{2\pi}\delta\omega^{\mathsf{x}}$ {[}kHz{]}\tabularnewline
 &  &  &  &  &  &  &  & \tabularnewline
\hline 
 &  &  &  &  &  &  &  & \tabularnewline
8 &  & $6000$ &  & $0.056$ &  & $0.18$ &  & $9.2$\tabularnewline
 &  &  &  &  &  &  &  & \tabularnewline
\hline 
\end{tabular}
\caption{The characteristics of the ion trap and lasers relevant for the examples studied in Sec.~\ref{sec:schwinger-example} and the associated gate parameters in Tables~\ref{tab:gate-I-schwinger} and \ref{tab:gate-II-schwinger}.}
\label{tab:trap-schwinger}
\end{table}

\end{document}